\documentclass[onecolumn,preprint]{aastex63}
\usepackage{xcolor}

\usepackage{natbib}
\usepackage{amsmath}
\usepackage{enumitem}

\usepackage{soul}
\setstcolor{red}

\shortauthors{Bernardinelli et al.}

\defcitealias{Bernstein2000}{BK}

\newcommand{\ie}{\textit{i.e.}}
\newcommand{\eg}{\textit{e.g.}}
\newcommand{\E}{\mathrm{E}}

\newcommand{\des}{\textit{DES}}

\newcommand{\eqq}[1]{Equation~(\ref{#1})}

\newcommand{\comment}[1]{}

\begin{document}

\title{Photometry of outer Solar System objects from the Dark Energy Survey I: photometric methods, light curve distributions and trans-Neptunian binaries}

\author[0000-0003-0743-9422]{Pedro H. Bernardinelli}
\altaffiliation{DiRAC Postdoctoral Fellow}
\affiliation{DiRAC Institute, Department of Astronomy, University of Washington, 3910 15th Ave NE, Seattle, WA, 98195, USA}
\email{phbern@uw.edu}

\author[0000-0002-8613-8259]{Gary M. Bernstein}
\affiliation{Department of Physics and Astronomy, University of Pennsylvania, Philadelphia, PA 19104, USA}
\email{garyb@upenn.edu}

\author[0000-0002-2092-3545]{Nicholas Jindal}
\affiliation{Department of Physics and Astronomy, University of Pennsylvania, Philadelphia, PA 19104, USA}
\affiliation{Department of Physics, Ohio State University, Columbus, OH, 43210, USA}

\author{T.~M.~C.~Abbott}
\affiliation{Cerro Tololo Inter-American Observatory, NSF's National Optical-Infrared Astronomy Research Laboratory, Casilla 603, La Serena, Chile}
\author{M.~Aguena}
\affiliation{Laborat\'orio Interinstitucional de e-Astronomia - LIneA, Rua Gal. Jos\'e Cristino 77, Rio de Janeiro, RJ - 20921-400, Brazil}
\author{F.~Andrade-Oliveira}
\affiliation{Department of Physics, University of Michigan, Ann Arbor, MI 48109, USA}
\author{J.~Annis}
\affiliation{Fermi National Accelerator Laboratory, P. O. Box 500, Batavia, IL 60510, USA}
\author{D.~Bacon}
\affiliation{Institute of Cosmology and Gravitation, University of Portsmouth, Portsmouth, PO1 3FX, UK}
\author{E.~Bertin}
\affiliation{CNRS, UMR 7095, Institut d'Astrophysique de Paris, F-75014, Paris, France}
\affiliation{Sorbonne Universit\'es, UPMC Univ Paris 06, UMR 7095, Institut d'Astrophysique de Paris, F-75014, Paris, France}
\author{D.~Brooks}
\affiliation{Department of Physics \& Astronomy, University College London, Gower Street, London, WC1E 6BT, UK}
\author{D.~L.~Burke}
\affiliation{Kavli Institute for Particle Astrophysics \& Cosmology, P. O. Box 2450, Stanford University, Stanford, CA 94305, USA}
\affiliation{SLAC National Accelerator Laboratory, Menlo Park, CA 94025, USA}
\author{A.~Carnero~Rosell}
\affiliation{Instituto de Astrofisica de Canarias, E-38205 La Laguna, Tenerife, Spain}
\affiliation{Laborat\'orio Interinstitucional de e-Astronomia - LIneA, Rua Gal. Jos\'e Cristino 77, Rio de Janeiro, RJ - 20921-400, Brazil}
\affiliation{Universidad de La Laguna, Dpto. Astrofísica, E-38206 La Laguna, Tenerife, Spain}
\author{M.~Carrasco~Kind}
\affiliation{Center for Astrophysical Surveys, National Center for Supercomputing Applications, 1205 West Clark St., Urbana, IL 61801, USA}
\affiliation{Department of Astronomy, University of Illinois at Urbana-Champaign, 1002 W. Green Street, Urbana, IL 61801, USA}
\author{J.~Carretero}
\affiliation{Institut de F\'{\i}sica d'Altes Energies (IFAE), The Barcelona Institute of Science and Technology, Campus UAB, 08193 Bellaterra (Barcelona) Spain}
\author{L.~N.~da Costa}
\affiliation{Laborat\'orio Interinstitucional de e-Astronomia - LIneA, Rua Gal. Jos\'e Cristino 77, Rio de Janeiro, RJ - 20921-400, Brazil}
\author{M.~E.~S.~Pereira}
\affiliation{Hamburger Sternwarte, Universit\"{a}t Hamburg, Gojenbergsweg 112, 21029 Hamburg, Germany}
\author{T.~M.~Davis}
\affiliation{School of Mathematics and Physics, University of Queensland,  Brisbane, QLD 4072, Australia}
\author{S.~Desai}
\affiliation{Department of Physics, IIT Hyderabad, Kandi, Telangana 502285, India}
\author{H.~T.~Diehl}
\affiliation{Fermi National Accelerator Laboratory, P. O. Box 500, Batavia, IL 60510, USA}
\author{P.~Doel}
\affiliation{Department of Physics \& Astronomy, University College London, Gower Street, London, WC1E 6BT, UK}
\author{S.~Everett}
\affiliation{Jet Propulsion Laboratory, California Institute of Technology, 4800 Oak Grove Dr., Pasadena, CA 91109, USA}
\author{I.~Ferrero}
\affiliation{Institute of Theoretical Astrophysics, University of Oslo. P.O. Box 1029 Blindern, NO-0315 Oslo, Norway}
\author{D.~Friedel}
\affiliation{Center for Astrophysical Surveys, National Center for Supercomputing Applications, 1205 West Clark St., Urbana, IL 61801, USA}
\author{J.~Frieman}
\affiliation{Fermi National Accelerator Laboratory, P. O. Box 500, Batavia, IL 60510, USA}
\affiliation{Kavli Institute for Cosmological Physics, University of Chicago, Chicago, IL 60637, USA}
\author{J.~Garc\'ia-Bellido}
\affiliation{Instituto de Fisica Teorica UAM/CSIC, Universidad Autonoma de Madrid, 28049 Madrid, Spain}
\author{G.~Giannini}
\affiliation{Institut de F\'{\i}sica d'Altes Energies (IFAE), The Barcelona Institute of Science and Technology, Campus UAB, 08193 Bellaterra (Barcelona) Spain}
\author{D.~Gruen}
\affiliation{University Observatory, Faculty of Physics, Ludwig-Maximilians-Universit\"at, Scheinerstr. 1, 81679 Munich, Germany}
\author{K.~Herner}
\affiliation{Fermi National Accelerator Laboratory, P. O. Box 500, Batavia, IL 60510, USA}
\author{S.~R.~Hinton}
\affiliation{School of Mathematics and Physics, University of Queensland,  Brisbane, QLD 4072, Australia}
\author{D.~L.~Hollowood}
\affiliation{Santa Cruz Institute for Particle Physics, Santa Cruz, CA 95064, USA}
\author{K.~Honscheid}
\affiliation{Center for Cosmology and Astro-Particle Physics, The Ohio State University, Columbus, OH 43210, USA}
\affiliation{Department of Physics, The Ohio State University, Columbus, OH 43210, USA}
\author{D.~J.~James}
\affiliation{Center for Astrophysics $\vert$ Harvard \& Smithsonian, 60 Garden Street, Cambridge, MA 02138, USA}
\author{K.~Kuehn}
\affiliation{Australian Astronomical Optics, Macquarie University, North Ryde, NSW 2113, Australia}
\affiliation{Lowell Observatory, 1400 Mars Hill Rd, Flagstaff, AZ 86001, USA}
\author{J. Mena-Fern{\'a}ndez}
\affiliation{Centro de Investigaciones Energ\'eticas, Medioambientales y Tecnol\'ogicas (CIEMAT), Madrid, Spain}
\author{F.~Menanteau}
\affiliation{Center for Astrophysical Surveys, National Center for Supercomputing Applications, 1205 West Clark St., Urbana, IL 61801, USA}
\affiliation{Department of Astronomy, University of Illinois at Urbana-Champaign, 1002 W. Green Street, Urbana, IL 61801, USA}
\author{R.~Miquel}
\affiliation{Instituci\'o Catalana de Recerca i Estudis Avan\c{c}ats, E-08010 Barcelona, Spain}
\affiliation{Institut de F\'{\i}sica d'Altes Energies (IFAE), The Barcelona Institute of Science and Technology, Campus UAB, 08193 Bellaterra (Barcelona) Spain}
\author{R.~L.~C.~Ogando}
\affiliation{Observat\'orio Nacional, Rua Gal. Jos\'e Cristino 77, Rio de Janeiro, RJ - 20921-400, Brazil}
\author{A.~Pieres}
\affiliation{Laborat\'orio Interinstitucional de e-Astronomia - LIneA, Rua Gal. Jos\'e Cristino 77, Rio de Janeiro, RJ - 20921-400, Brazil}
\affiliation{Observat\'orio Nacional, Rua Gal. Jos\'e Cristino 77, Rio de Janeiro, RJ - 20921-400, Brazil}
\author{A.~A.~Plazas~Malag\'on}
\affiliation{Department of Astrophysical Sciences, Princeton University, Peyton Hall, Princeton, NJ 08544, USA}
\author{M.~Raveri}
\affiliation{Department of Physics, University of Genova and INFN, Via Dodecaneso 33, 16146, Genova, Italy}
\author{E.~Sanchez}
\affiliation{Centro de Investigaciones Energ\'eticas, Medioambientales y Tecnol\'ogicas (CIEMAT), Madrid, Spain}
\author{I.~Sevilla-Noarbe}
\affiliation{Centro de Investigaciones Energ\'eticas, Medioambientales y Tecnol\'ogicas (CIEMAT), Madrid, Spain}
\author{M.~Smith}
\affiliation{School of Physics and Astronomy, University of Southampton,  Southampton, SO17 1BJ, UK}
\author{E.~Suchyta}
\affiliation{Computer Science and Mathematics Division, Oak Ridge National Laboratory, Oak Ridge, TN 37831}
%\author{M.~E.~C.~Swanson}
%\affiliation{}
\author{G.~Tarle}
\affiliation{Department of Physics, University of Michigan, Ann Arbor, MI 48109, USA}
\author{C.~To}
\affiliation{Center for Cosmology and Astro-Particle Physics, The Ohio State University, Columbus, OH 43210, USA}
\author{A.~R.~Walker}
\affiliation{Cerro Tololo Inter-American Observatory, NSF's National Optical-Infrared Astronomy Research Laboratory, Casilla 603, La Serena, Chile}
\author{P.~Wiseman}
\affiliation{School of Physics and Astronomy, University of Southampton,  Southampton, SO17 1BJ, UK}
\author{Y.~Zhang}
\affiliation{Cerro Tololo Inter-American Observatory, NSF's National Optical-Infrared Astronomy Research Laboratory, Casilla 603, La Serena, Chile}
\affiliation{Department of Astronomy, University of Michigan, Ann Arbor, MI 48109, USA}

\collaboration{1000}{(The \des\ Collaboration)}
\suppressAffiliations

\begin{abstract}
We report the methods of and initial scientific inferences from the extraction of precision photometric information for the $>800$ trans-Neptunian objects (TNOs) discovered in the images of the \textit{Dark Energy Survey (DES).}  Scene-modelling photometry is used to obtain shot-noise-limited flux measures for each exposure of each TNO, with background sources subtracted.  Comparison of double-source fits to the pixel data with single-source fits are used to {identify and characterize two binary TNO systems}. A Markov Chain Monte Carlo method samples the joint likelihood of the intrinsic colors of each source as well as the amplitude of its flux variation, given the time series of multiband flux measurements and their uncertainties.  A catalog of these colors and light curve amplitudes $A$ is included with this publication.  We show how to assign a likelihood to the distribution $q(A)$ of light curve amplitudes in any subpopulation.  Using this method, we find decisive evidence {(\ie\ evidence ratio $<0.01$)} that cold classical (CC) TNOs with absolute magnitude $6<H_r<8.2$ are more variable than the hot classical (HC) population of the same $H_r$, reinforcing theories that the former form \textit{in situ} and the latter arise from a different physical population. 
Resonant and scattering TNOs in this $H_r$ range have variability consistent with either the HC's or CC's.  \des\ TNOs with 
$H_r<6$ are seen to be decisively less variable than higher-$H_r$ members of any dynamical group, as expected.  More surprising is that detached TNOs are decisively less variable than scattering TNOs, which requires them to have distinct source regions or some subsequent differential processing.
\end{abstract}
\reportnum{}


\section{Introduction} \label{sec:intro}

The trans-Neptunian region is a distant reservoir of small bodies that trace the formation history of the Solar System \citep{Nesvorny2018a}. We currently know of more than 3000 of these objects, with recent surveys capable of discovering several hundreds at a time \citep[\emph{e.g.}][]{Petit2011,Bannister2018,Bernardinelli2022}. The combination of dynamical and physical characterizations of these populations has led to our understanding of several key aspects of the formation of the outer Solar System \citep[see][for a recent review]{Gladman2021}. Photometric measurements of these trans-Neptunian objects (TNOs) are of particular interest, and analyses of such data have led to the understanding of the bulk properties of surface shapes \citep{Showalter2021}, the determination of distinct compositional classes from surface colors \citep[\emph{e.g.}][]{2008ssbn.book...91D,Fraser2012,Schwamb2019}, characterization of the size distribution \citep[\emph{e.g.}][]{Bernstein2004,Fraser2014,Kavelaars2021}, discovery of a large fraction of binary systems \citep{Stephens2006,Parker2011a,2020tnss.book..201N}, and the identification of a collisional family \citep{Brown2007}.

The Dark Energy Survey \citep[DES,][]{TheDarkEnergySurveyCollaboration2005a,Abbott2016} received an allocation of 575 nights on the 4m Blanco telescope at Cerro Tololo, using the Dark Energy Camera \citep[DECam,][]{Flaugher2015} to cover $5000\deg^2$ of the sky in the $grizY$ photometric system from 2013 to 2019. The survey's primary objective has been to study the distribution of dark matter and the nature of dark energy \citep{DESCollaboration2022}, but the data have enabled the discovery of hundreds of outer Solar System objects \citep{Bernardinelli2019,Bernardinelli2022} - we refer the reader to these two publications for a comprehensive presentation of the discovery pipeline. All 814 \des\ objects have been dynamically classified following \cite{Khain2020}, and the DES survey simulator \citep{Bernardinelli2022} allows us to carefully estimate our detection biases as a function of dynamical population, magnitude, color and light curve amplitude.

{The \des\ observing strategy was optimized for extragalactic science, which means that it is less efficient 
in terms of TNO discoveries per night of telescope time than surveys designed for TNO discovery \citep{Bannister2018,DEEPI}.} \des\ observes a given region too many times, in too many filters, over too long a time span, to be optimal for discovery.  This redundancy has the advantage that we can extract significantly more information about each discovered source than an optimized discovery search.  Each object has been observed many times: between 6-10 times in each of the $grizY$ bands, on average, over the six years of data collection. This means that each object is imaged on average between 30 and 50 times, depending on whether or not the object was inside the footprint for the entirety of the survey and each region's cadence.  This allows estimation of colors in the $griz$ bands ($Y$ is generally too low in signal-to-noise ratio, SNR) and also an estimate of the time variability of each source.  Inferring colors and variability simultaneously allows the color estimates to include uncertainties that arise when colors are measured from non-simultaneous exposures in the presence of variability, while also exploiting the occasions when \des\ targets a given source in multiple filters in quick succession.  This paper will present the methods and results of such estimates for the full \des\ TNO catalog, as well as some initial physical inferences that can be made from the variability information.  Section~\ref{sec:smp} will describe the extraction of fluxes, and Section~\ref{sec:colormc} the extraction of colors and light curve amplitudes (LCA's) from the flux time series.  In Section~\ref{sec:lca}  we examine the distributions of light-curve amplitudes in different TNO sub-populations as an indicator of different physical states.  Scientific analyses of the color catalog will appear in future publications.

The process of extracting optimal flux estimates from the pixel data can also be generalized to fit two fluxes of a potential binary pair to each exposure of a given TNO.  In Section~\ref{sec:binarysmp} we describe this process and present the resulting binary candidates found among the \des\ TNOs.

Accompanying this publication is a data release with the $\approx 30,000$ photometric measurements, absolute magnitudes, colors and light curve amplitudes of our objects, as well as some of the software required for the analysis we present here. The data release is available at [forthcoming], and discussed in Section \ref{sec:datarelease}.

\section{Individual photometric measurements}
\label{sec:smp}
To extract unbiased flux measurements of each TNO from the \des\ images, we use the images in which the object is detected and also images the orbit predicts should contain the object, but there is no detection in the catalog (``non-detections'')---similar to the sub-threshold significance (\texttt{STS}) measurement of \cite{Bernardinelli2019}.   We will collectively call these ``observations.''
We first start by describing the photometry for each individual observation of a TNO.

We determine the TNO positions in each exposure by using the values predicted by the orbit fit, and model each detection using the scene modeling photometry (SMP) technique commonly used in supernovae type Ia cosmology \citep{Brout2019}. We use the DES point spread function (PSF) model of \cite{Jarvis2020}, with PSFs derived for the full field of view of each DES exposure. A full description of the \des\ calibration procedure is presented in \cite{Burke2017} and \cite{Sevilla-Noarbe2020}, and comparisons between \des\ and Gaia show an exquisite calibration with 3 mmag root-mean-square differences between the two surveys \citep{DESDR2}.

Each location where a TNO is observed {on some single night} has $n-1$ images in the same band at different epochs in which the TNO is \emph{not} present. We posit that the background is composed of point sources on a $m \times m$ {square grid, with spacing of roughly the $1\sigma$ width of the PSF, so the PSF will blur them into a smooth distribution to represent any extended sources.  The grid sources have fluxes $\mathbf{P}_{uv},$ and are} centered on the TNO location. {Given the $\sim0.95\arcsec$ FWHM of typical \des\ imaging, we} place sources every $0.35\arcsec$ and define $m = 20$, so these sources span a $7\arcsec \times 7\arcsec$ region. These point sources are then mapped into the $(u,v)$ pixel coordinates of each image by inverting the DECam astrometric model \citep{Bernstein2017}.  The astrometric solution and the PSF models are both functions of source color; we begin by assuming a nominal $g - i = 0.61$ color{ (typical of stellar sources)} for each {background} source.  We will let the $u,v$ symbols serve both as indices into the grid of background sources, as well as their exact positions in the pixel coordinate system.
The expected signal in pixel $(i,j)$ for exposure $\mu$ for this mosaic of sources is modelled as the convolution of each $\mathbf{P}_{uv}$ with exposure's PSF derived for its location $(u,v);$ plus some constant background level $\mathbf{b}_\mu$:
\begin{equation}
	\mathbf{M}^\mu_{ij} = \sum_{u,v} \mathbf{PSF}_\mu(i-u,j-v) \mathbf{P}_{uv} + \mathbf{b}_\mu.
      \end{equation}
In the {single} exposure $\nu$ where the TNO is present, we adopt the same model with an additional point-source term:

\begin{equation}
	\mathbf{M}^\nu_{ij} = \sum_{u,v} \mathbf{PSF}_\nu(i-u,j-v) \mathbf{P}_{uv} + \mathbf{b}_\nu + f_\mathrm{TNO} \mathbf{PSF}_\nu(i,j). \label{eq:model}
\end{equation}
That is, $f_\mathrm{TNO}$ represents the integrated flux of the TNO {at this epoch and band}. Due to the short exposures times (90 seconds for most images), these sources are not trailed, and so corrections such as pill apertures \citep[\emph{e.g.}][]{Fraser2016} are not needed.  {Initially we use a default color for the TNO in evaluating the PSF and its expected pixel position.}

This model, then, has $N = n + m^2 + 1$ free parameters, and $nk^2 - N$ degrees of freedom, where $k>m$ is the number of pixels in the postage stamp modeled (we used stamps with $30 \times 30$ pixels, corresponding to a sky area of $7.8\arcsec \times 7.8\arcsec$). We fit these using a least squares minimization, comparing the model to the measured pixel values $\mathbf{Im}_\mu$:
\begin{equation}
	\chi^2 = \sum_{i,j} \frac{(k_\nu\mathbf{Im}^{\nu}_{ij}  - \mathbf{M}^{\nu}_{ij})^2}{\sigma^2_{\nu,ij}} +  \sum_{\mu, ij} \frac{(k_\mu \mathbf{Im}^{\mu}_{i,j}  - \mathbf{M}^{\mu}_{ij})^2}{\sigma^2_{\mu,ij}}  . \label{eq:lsq}
\end{equation}
The (constant) $k_{\mu,\nu}$ terms correct for the different zero-points in each exposure, bringing them to a common flux scale \citep{Burke2017}, and $\sigma^2_{\mu,ij}$ is the noise variance at each pixel from sky background and detector noise. This model is a linear system determined by a design matrix $\mathbf{A}$ that contains all PSF realizations and the constant, unitless background terms; parameters $\mathbf{X} = \{\mathbf{P}_{uv}, \mathbf{b}_\mu, f_\mathrm{TNO}\}$; and the target matrix $\mathbf{Y}$ (i.e., $\mathbf{Y} = \mathbf{A} \mathbf{X}$) that represents all images. Thus, we can solve for the parameters using a standard linear least-squares solution. 

\begin{figure}[ht!]
	\centering
	\includegraphics[width=0.8\textwidth]{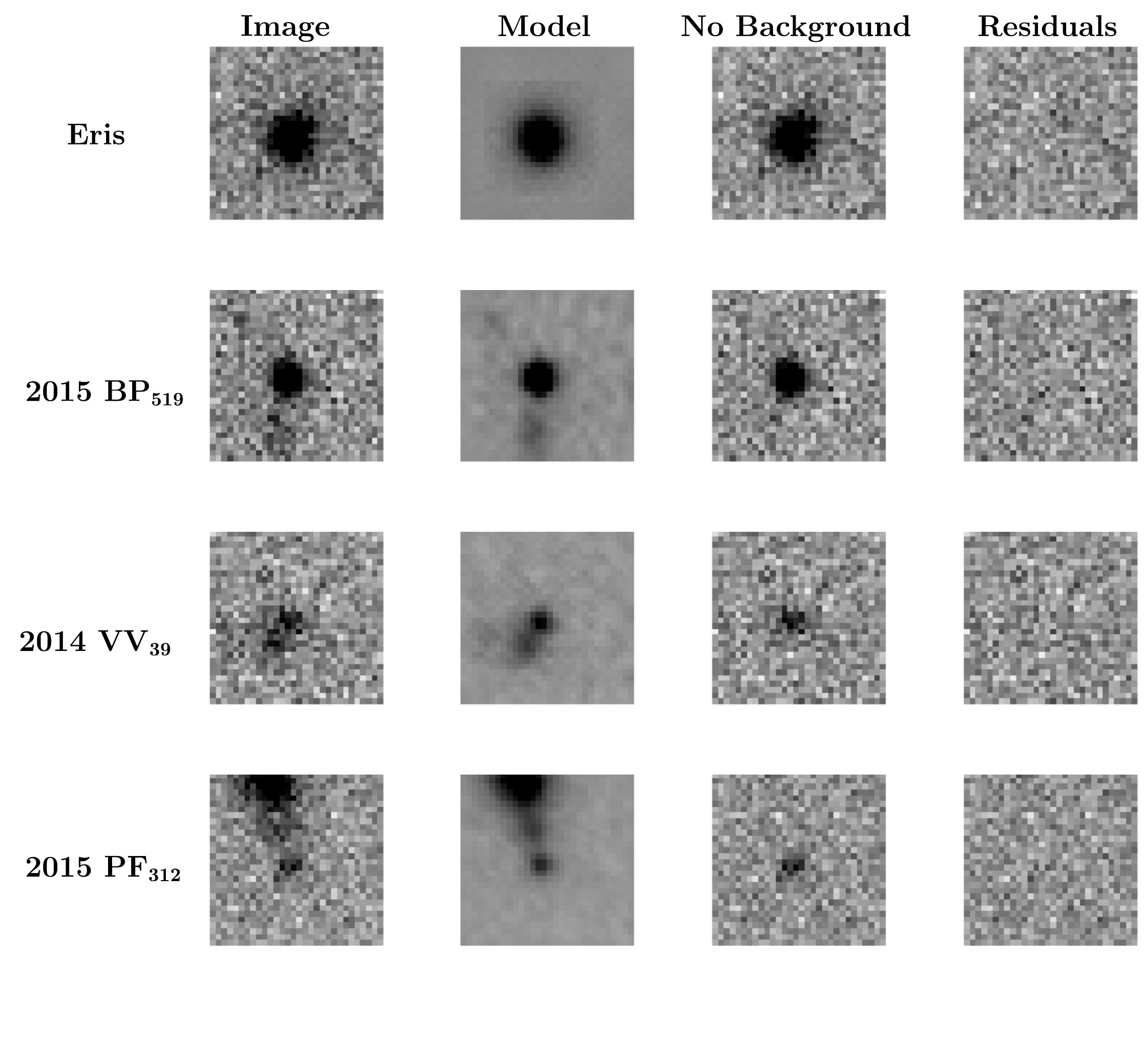}
	\caption{Examples of successful scene-modelling measurements for a few objects. The columns, from left to right for each object are: 1) a postage stamp of the data centered on the detection; 2) the best-fit model in Equation \ref{eq:model}; 3) the difference between the data and the background portion of the model (i.e. without subtracting the model of the TNO itself) and 4) the residuals of data minus the full model. The first row shows an example measurement of our brightest object, the magnitude $m_r \approx 19.5$ Eris.\label{im:goodsmp}}
\end{figure}

\begin{figure}[ht!]
	\centering
	\includegraphics[width=0.8\textwidth]{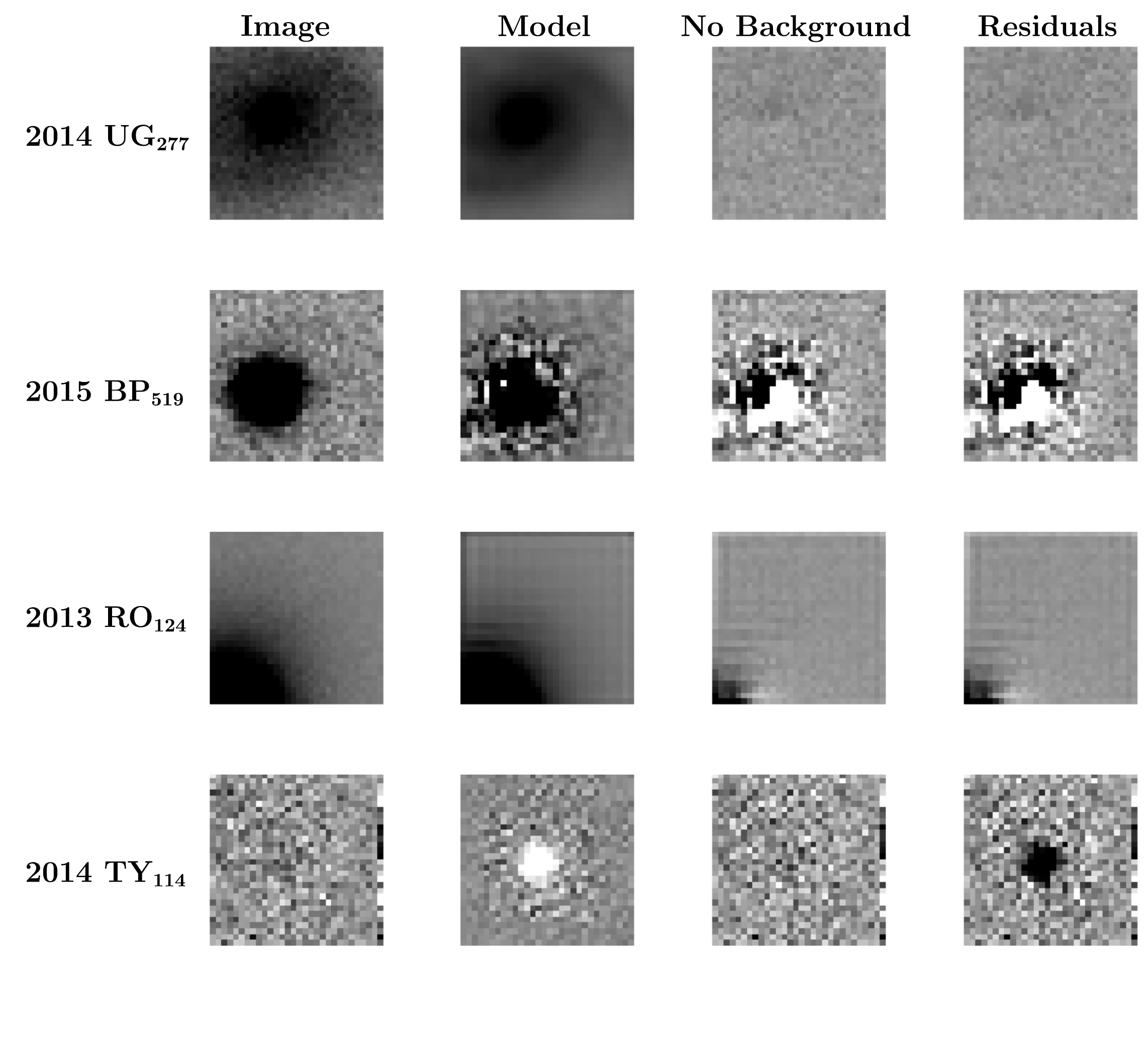}
	\caption{Same columns as in Figure \ref{im:goodsmp}, but the four examples show failures of the scene-modelling procedure. The first row shows a case where the TNO was near an extended background galaxy. This example is remarkable because, even though there is no usable photometry for the TNO, the model reproduces the spiral structure of the galaxy really well, leading to a near perfect subtraction. The second row shows a similar case of a bright background source near the TNO, but in this case the model failed to reproduce this background. The third row corresponds to a bright source near the edge of the stamp, with its scattered light precluding any photometric measurement nearby. The final row is a case where the least squares procedure of Equation \ref{eq:lsq} led to an incorrect solution (namely, the model predicts a large negative flux for the central point source).\label{im:badsmp}}
\end{figure}

{The initial estimates of the $\sigma_{\nu,ij}$ account only for shot noise from the uniform sky background, and for detector noise.}
To account for the additional variance due to shot noise from the sources, we update the weights in exposure $\nu$ using the first least-squares solution for $\mathbf{M}^\nu_{ij}$: 
\begin{equation}
	\sigma_{\nu,ij}^2 \to \sigma_{\nu,ij}^2 + \frac{\max(\mathbf{M}^\nu_{ij},0)}{g_{\nu,ij}},
\end{equation}
where $g_{\nu,ij}$ is the amplifier-dependent gain in each pixel and exposure \citep{Bernstein2018}. We refit the model with these new variances, and derive the flux uncertainties from the photometric solution.  {Once all of the exposures of a given TNO have been measured, we estimate its mean $g-i$ color,\footnote{{All \des\ PSF and photometric color dependence is parameterized by $g-i$ of a stellar source.}}  then repeat the entire measurement process for each TNO apparition while using this $g-i$ value in the color-dependent astrometric solution and PSF for the TNO flux.}
All $\approx$30,000 individual observations were visually inspected, and cases where the scene-modelling procedure failed were discarded. We show successful examples in Figure \ref{im:goodsmp} and failures in Figure \ref{im:badsmp}.

We note that this methodology can be easily applied to extended sources (for example, comets) and to larger areas of the sky: in \cite{BB2021} we applied this methodology to $400 \times 400$ pixel stamps in order to detect C/2014 UN$_{271}$'s extended coma. A generalization to binary sources is presented in the next section.

\section{Identifying binaries with scene-modelling photometry}
\label{sec:binarysmp}
\subsection{Methodology}
The modification for a binary TNO requires changing Equation \ref{eq:model} to include a second point source term:
\begin{equation}
	\mathbf{M}^\nu_{ij} = \sum_{u,v} \mathbf{PSF}_\nu(i-u,j-v) \mathbf{P}_{uv} + \mathbf{b}_\nu + f_1 \mathbf{PSF}_\nu(i + \Delta x_1,j + \Delta y_1) + f_2 \mathbf{PSF}_\nu (i + \Delta x_2, j + \Delta y_2), \label{eq:binary}
\end{equation}
where the indices 1 and 2 correspond to the primary and secondary sources, respectively. For a fixed set of $\{ \Delta x_1, \Delta y_1, \Delta x_2, \Delta y_2\}$, the linearity of the least squares procedure in Equation \ref{eq:lsq} is preserved\footnote{Note that this is also the case for a single source with a free centroid, and so the same methodology can be used to adjust the location of the PSF. This is particularly useful for bright sources, where $\approx 10$ mas shifts in the center can account for uncertainties due to atmospheric turbulence.}. The minimization of the $\chi^2$ of the binary model proceeds by first obtaining the shifted PSFs, and then fitting for $\{f_1, f_2\}$ by solving the linear least-squares procedure as in Equation \ref{eq:lsq}.

\begin{figure}[ht!]
	\centering
	\includegraphics[width=0.8\textwidth]{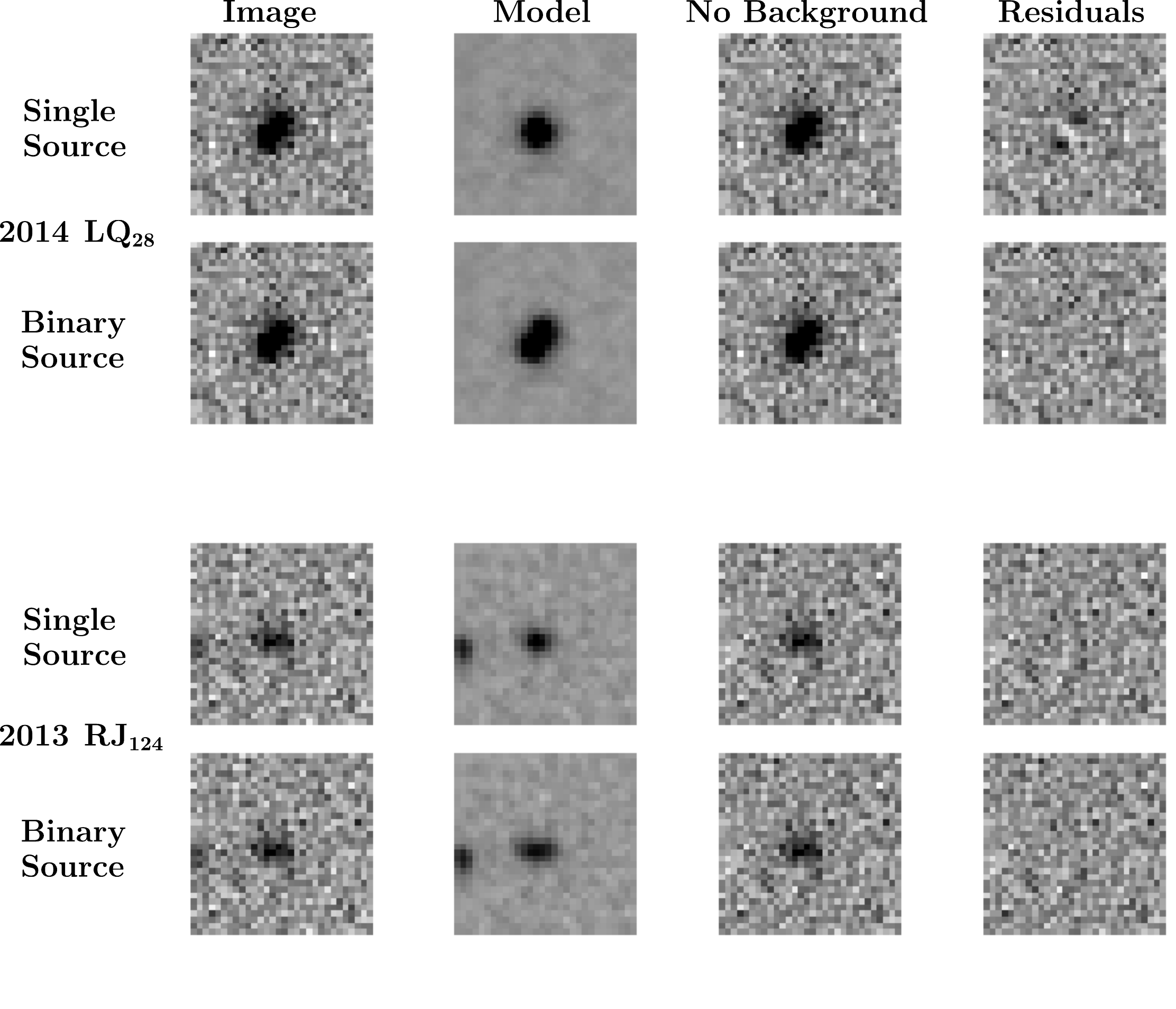}
	\caption{Single (upper row of each object) and binary (bottom row) point source scene-modelling photometry applied to the same detections of two high-confidence binary TNOs identified in the \des\ data. In the first case ($\Delta \chi^2 = 130.6$), a distinct quadrupole pattern is present in the residual image, a strong indication that this source is a binary. The second object ($\Delta\chi^2 = 12.8$), while less visually apparent, is statistically significant across several images. \label{im:binary}}
\end{figure}

To determine whether the binary model is preferable to the single PSF model, we check the difference in $\chi^2$ between the two, $\Delta \chi^2 = \chi^2_{\mathrm{single}} - \chi^2_\mathrm{binary}$. As we have an additional 5 parameters (the flux of the secondary source and the position shifts), we expect $\Delta\chi^2$ to also follow a $\chi^2$ distribution with 5 degrees of freedom. However, we note that there are other reasons (besides a binary object) for the $\chi^2$ to improve, such as a poorly modeled background feature (such as a star), a miscentering in the PSF of the main source (as the position is fixed, and derived from the orbit fitting in our single PSF photometry), or \emph{another} transient in the same image (such as an asteroid). We visually inspect all cases where $\Delta\chi^2 \geq 9$, where the probability of the null hypothesis being true (i.e., a single source) is $0.1\%$. Figure \ref{im:binary} shows two examples for the high confidence binaries identified in this data. 

\begin{figure}[ht!]
	\centering
	\includegraphics[width=0.8\textwidth]{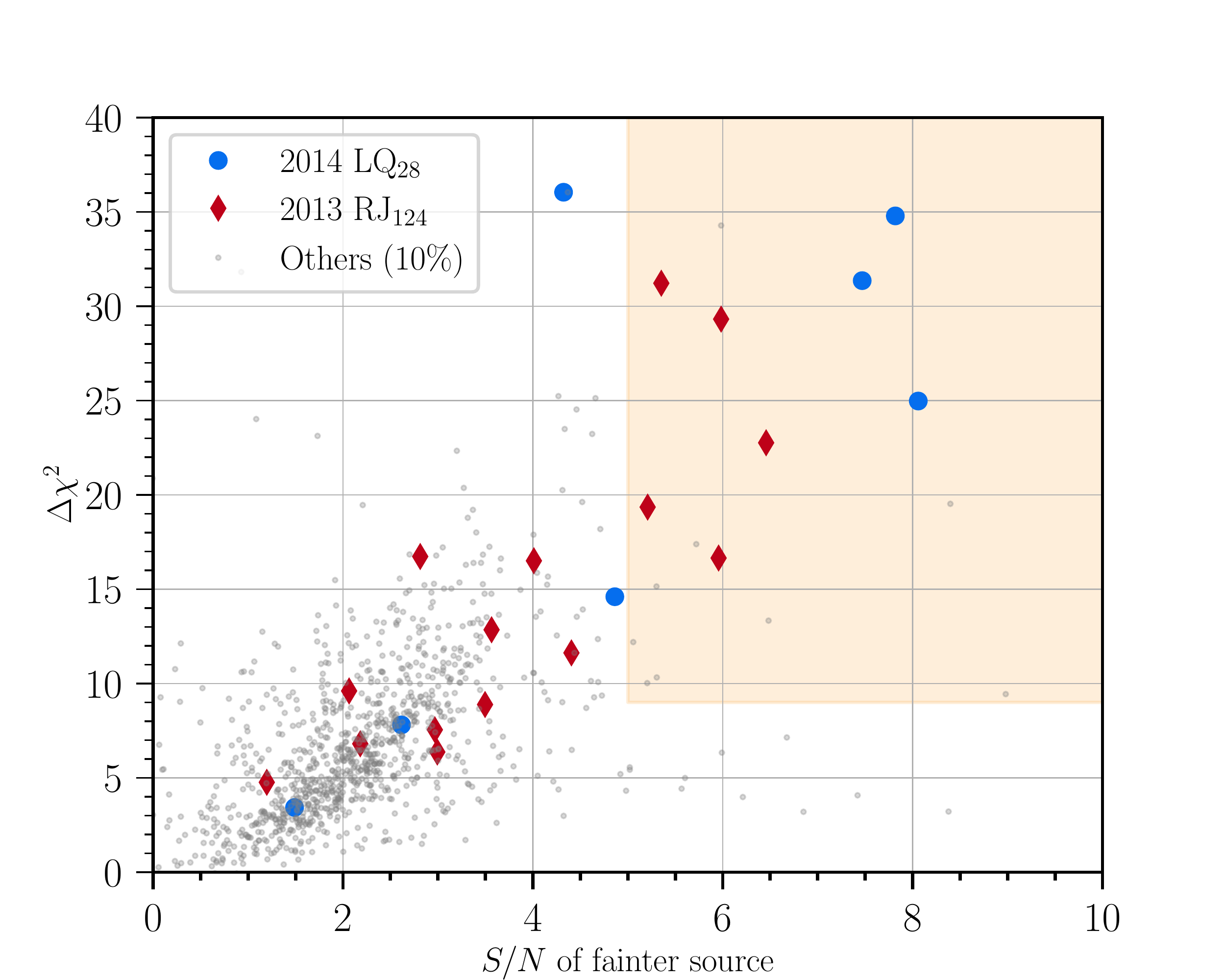}
	\caption{$\Delta \chi^2$ between the single source and binary models vs $S/N$ of the fainter source for a randomly chosen subset of 10\% of the $\approx25,000$ photometric measurements (in gray) in the DES data. The two binary objects are shown with different colors, and the acceptance region is shaded in yellow. The other cases of objects where the binary fit yields a statistically significant improvement in the $\chi^2$ {arise when there is a poorly subtracted background object in the single-source fit, or when the binary fitting corrects small errors in the assumed pixel position of the primary object caused by atmospheric turbulence.}
\label{im:deltachi2vssn}}
\end{figure}

The application of this technique to all our $\approx 25,000$ detections in the $griz$ bands leads to 2 objects where several images led to both an improvement in the $\chi^2,$ and $S/N \geq 5$ for the fainter source in multiple images, as shown in Figure \ref{im:deltachi2vssn}. We also note that Eris's satellite Dysnomia is \emph{not} resolved in the \des\ images \citep{bernstein2023}. While 2014 LQ$_{28}$ had been previously identified as a binary \citep{thirouin2019}, 2013 RJ$_{124}$ is a new binary discovery. The objects (612620) 2003 SQ$_{317}$, 2014 QL$_{441}$ and 2016 TT$_{94}$ also show two images each with $\Delta \chi^2>10$ and $S/N \geq 5$ with no associated transient or artifact, but these images are not enough to confirm their binary status, and require follow-up observations. We note that 2003 SQ$_{317}$ has been imaged as a single point source with HST in \cite{2007hst..prop11113N}, which indicates a probable false positive from our analysis.

\subsection{Mutual orbit determination and derived properties}
Our two binaries have multiple resolved observations across several years of the survey, we can attempt to determine a mutual orbit for these systems. In this particular case, we can relax our requirement for a resolved observation to $\Delta \chi^2 \geq 8$: this is justified, as now we have a strong ``prior'' that this is a binary object, and the probability of the null hypothesis is still low ($0.7\%$). We can determine the uncertainty in the shifts from the center of mass position by inverting the Hessian of the $\chi^2$ of the model given by Equation \ref{eq:binary}. 

These position shifts (and their corresponding uncertainties) can be readily transformed into a separation vector $r$ and position angle $\phi \in [0\degr,180\degr]$, avoiding any degeneracy in the determination of the primary source. We fit to each object a Keplerian orbit, where the 6 orbital parameters $(a_m,e_m,i_m,\Omega_m,\omega_m,\mathcal{M}_m)$ and the period $P_m$ are used to derive distance and light-travel time corrected sky-plane projections $(\hat{r},\hat{\phi})$. Here, the angles refer to the equatorial plane. Indexing the resolved images by $\mu$, we use a Markov Chain Monte Carlo to sample this seven dimensional parameter space with the likelihood 
\begin{equation}
	\mathcal{L} \propto \prod_{\mu} \exp\left[-\frac{\left( r_\mu - \hat{r}(t_\mu), \phi - \hat{\phi}_\mu \right) \Sigma_\mu^{-1} \left( r_\mu - \hat{r}(t_\mu) , \phi - \hat{\phi}_\mu \right) ^\top}{2}\right],
\end{equation}
 A uniform prior is also applied to restrict these parameters to physically reasonable values for a bound orbit ($a_m,P_m > 0$, $0 \leq \Omega_m, \omega_m, \mathcal{M}_m \leq 2 \pi$, $0 \leq i_m \leq \pi$, $0 \leq e_m \leq 1$). We present the results of these fits in Table \ref{tb:orbits}. In particular, we note that only our semi-major axes, eccentricities and periods are well constrained for each system, as indicated by the large permissible ranges of $(i_m, \Omega_m, \omega_m,\mathcal{M}_m)$. That is, these large error bars indicate that the orientation of the orbit is poorly constrained. From the orbital elements, we can also derive the masses of these systems (using Kepler's third law), as well as the ratio between the mutual semi-major axis and the system's Hill radius (see, \emph{e.g.}, \citealt{Parker2011a}). 

\begin{deluxetable}{l|ccccccccc|cc}
\tablewidth{0pt}
\tablecaption{Mutual orbits of the trans-Neptunian binaries}
\tabletypesize{\small}
  \tablehead{\colhead{Object} & \colhead{$a_m$} & \colhead{$e_m$} & \colhead{$i_m$} & \colhead{$\Omega_m$} & \colhead{$\omega_m$} & \colhead{$\mathcal{M}_m$} & \colhead{$P_m$ } & \colhead{Epoch} & \colhead{Mass} & \colhead{Hill radius}\\
  \colhead{} & \colhead{$10^6\,$m} &\colhead{} & \colhead{deg} & \colhead{deg} & \colhead{deg} & \colhead{deg} & \colhead{days} & \colhead{MJD} & \colhead{$10^{18}\,$kg} & \colhead{$10^7\,$m}}
\startdata
2014 LQ$_{28}$ & $32.9^{+7.1}_{-3.5}$ & $0.64^{+0.14}_{-0.16}$ & $43^{+2}_{-2}$ & $5^{+5}_{-3}$ & $256^{+4}_{-6}$ & $268^{+18}_{-20}$ & $1470^{+390}_{-230}$ &  57615.258 & $1.4^{+0.7}_{-0.5}$ &  $52^{+7}_{-7}$\\
2013 RJ$_{124}$ & $34.9^{+6.6}_{-4.0}$ & $0.14^{+0.13}_{-0.09}$ & $90^{+14}_{-15}$ & $103^{+49}_{-40}$ & $152^{+162}_{-43}$ & $38^{+121}_{-22}$ & $1850^{+290}_{-250}$ & 56545.355 & $1.0^{+0.3}_{-0.2}$  & $49^{+5}_{-4}$ \\
\enddata
\tablecomments{All values correspond to the 68\% limits of the posterior distribution marginalized over all other parameters. }
\label{tb:orbits}
\end{deluxetable}

In additional to astrometric data, we can also determine flux ratios. We define the magnitude difference $\delta m \equiv |2.5 \log_{10}(f_1/f_2)|$, where we take the absolute value to avoid ambiguity in determining which object is the primary in the PSF fitting procedure. We also have a few back-to-back observations (where the orientation of the system does not change) in multiple bands, and we can immediately determine colors without any ambiguity. These results are included in the data release (see Section \ref{sec:datarelease}). 

\begin{figure}[ht!]
	\centering
	\includegraphics[width=0.8\textwidth]{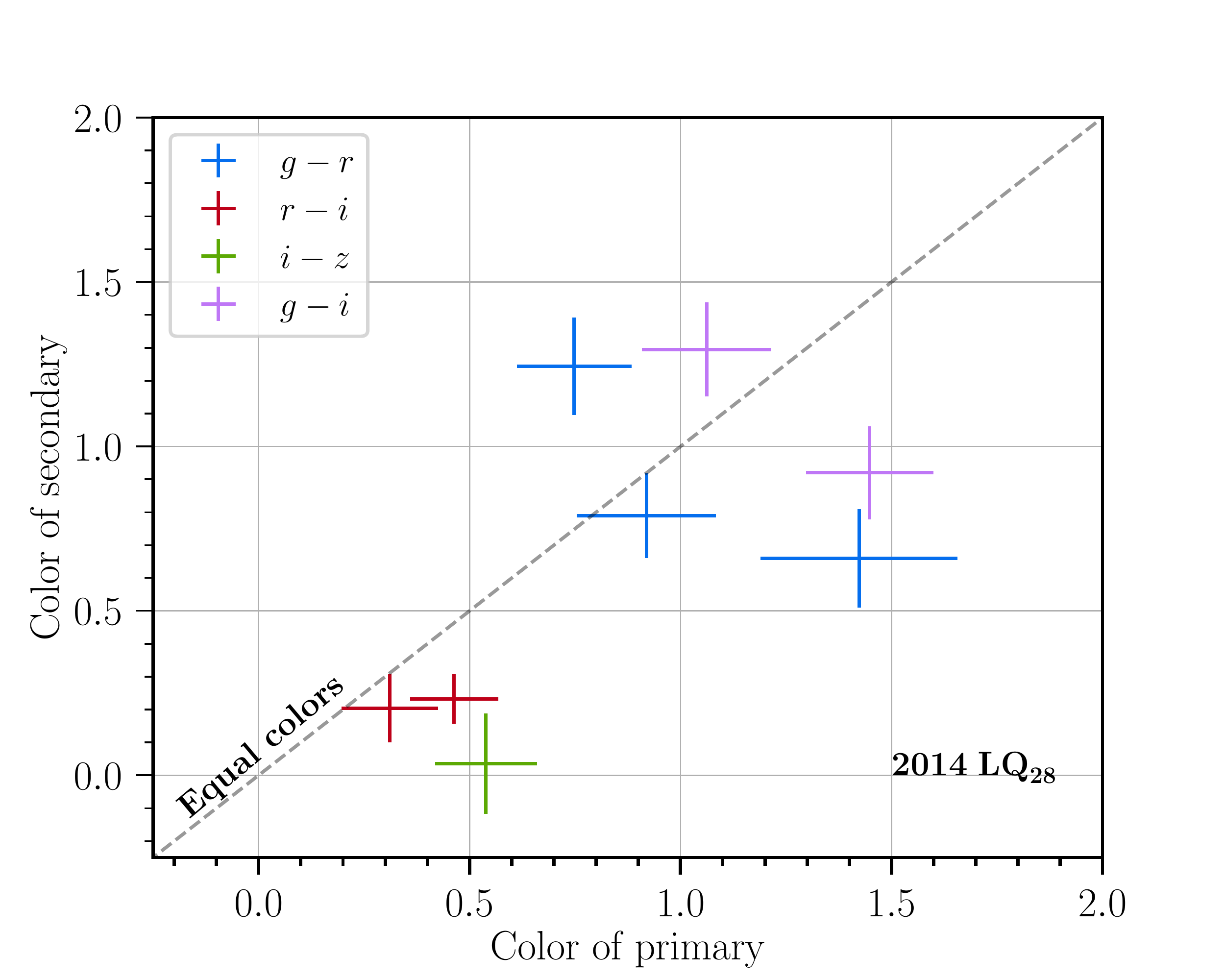}
	\vspace{-0.1cm}
	\caption{$g-r$ (blue), $r-i$ (red), $i-z$ (green) and $g-i$ (purple) colors and corresponding uncertainties measured from sequential exposures of each component of the 2014 LQ$_{28}$ system. The dashed gray diagonal line corresponds to the regime where the colors are equal in both members. The colors scatter around this diagonal line, and are all within $3\sigma$ of the nominal identity, implying that each member has the same colors (and therefore, similar surface composition).\label{im:bincol}}
\end{figure}

2014 LQ$_{28}$ has several observation pairs, presented in Figure \ref{im:bincol}. All color pairs are less than $3\sigma$ away from the colors being equal, and so the results are consistent with \cite{Benecchi2009}, with the colors of the primary being statistically indistinguishable from the secondary, implying similarities in the surface composition of each member of the system. 

\section{Extracting colors and light curve amplitudes}
\label{sec:colormc}
After the scene-modelling photometry, each TNO $i\in\{1,2,\ldots,N_{\rm TNO}\}$ has a series of measured fluxes $f_{ij}$ for $j\in\{1,\ldots,N_{{\rm obs},i}\}$ observations in bands $b_{ij},$ with nearly Gaussian uncertainties $\sigma_{ij}$. We will assume here that the fluxes have been adjusted for their heliocentric and geocentric distances to represent fluxes that would be observed with both distances at $d_{\rm ref}=30$~AU. 
What are the best estimates of the mean fluxes $\bar f_{ib}$ for source $i$ in band $b$?  From these, the best estimated absolute magnitudes $H_{ib}$ and colors $(b-b^\prime)_i \equiv H_{ib}-H_{ib^\prime}$ can be determined, and their uncertainties. 

A key element of the uncertainties is the potential for variability in the TNO fluxes due to rotation coupled with asphericity and/or surface inhomogeneities of the sources.  The level of variability is a quantity of interest itself as an indicator of the physical state of the TNO, as well as being a source of noise in flux/color measures, so we would like to have a principled estimate (and uncertainty) of the variability amplitude, as well as estimates of mean flux/color that have been marginalized over the variability.

The traditional means to determining colors would be to obtain high-$S/N$ multiband observations of each source within a time interval that is short compared to the variability period, allowing the measure of ``instantaneous'' colors.  Alternatively one could take enough observations in each band to be able to determine the period, reconstruct the light curve as a function of variability phase $\phi,$ and get phase-averaged fluxes $\bar f_{ib}$ for each band $b$.  Unfortunately the \des\ observing cadence does not admit either method.  In each band, the source is typically observed 8 times over a five-year span, far too sparse to determine a period, never mind construct a light curve.  Furthermore only occasionally are two observations of a given TNO are made on the same night, making instantaneous colors usually unavailable or low-$S/N.$  Consecutive observations in distinct bands do occur, however, so we would like a method that can exploit these events for deriving accurate colors.

We have developed a method to make optimal use of the information that we do have in determining the mean fluxes and the level of variability. {Our approach is conceptually similar to that used by \citet{Schemel21} to measure photometric variability of Jovian Trojans.}  We  assume a model in which the flux at observation $j$ is assumed to equal
\begin{equation}
  \hat f_j = \bar f_{b_j} \left[ 1 + Ah(\phi_j)\right].
\end{equation}
Here we have dropped the index $i$ of the TNO, for brevity, since the process is independent for each TNO.  A phase function $h(\phi)$ is a function of the variability phase $\phi \in [0,1)$ with $\langle h \rangle = 0$ when averaged over phase, and $\max_\phi h - \min_\phi h = 2$.  The parameter $A$ then gives the semi-amplitude of the fractional variation of true flux, and $\bar f_b$ is the time-averaged mean flux in band $b$.  Note that we define the mean and the light-curve semi-amplitude (LCA) in flux space rather than in magnitude space.  The peak-to-peak magnitude variation would be $\Delta m \equiv 2.5\log_{10} (1+A)/(1-A).$  We assume the convention $A\ge0,$ and only $A<1$ is physically possible.  We make the following critical assumptions:
\begin{itemize}
\item All bands have the same phase function $h$ and amplitude $A,$ \ie\ the variation is achromatic. 
\item The variability phase $\phi_j$ is a random unit deviate for each observation, \ie\ we know nothing about the period except that it is short enough to leave us with no knowledge of the relative phases of different observations---\emph{unless} observations $j$ and $k$ are within 1~hour of each other, in which case we assume $\phi_j=\phi_k$ but the value is unknown.
\item We will assume that the phase function is $h(\phi)=\sin(2\pi\phi).$  This is likely inaccurate for non-ellipsoidal geometries, but our goal is to obtain $A$ values that are indicative of variability level, so as to admit comparison of populations' variabilities, so we don't care that individual values of $A$ are precise LCA's.  
\end{itemize}

With the first assumption in hand we can write the probability of the observations $f_j$ with Gaussian uncertainties $\sigma_j$ as
\begin{equation}
  p\left(\{f_j\} | \{\bar f_b\}, A, \{\phi_j\}\right) \propto \prod_j \exp -\frac{1}{2}\left(\frac{ f_j - \bar f_{b_j} \left[ 1 + Ah(\phi_j)\right]}{\sigma_j}\right)^2.
\label{pphi}
\end{equation}
To obtain $p\left(\{f_j\} | \{\bar f_b\}, A\right)$ we need to marginalize \eqq{pphi} over all possible phases using the second assumption.  We divide all the observations $j$ into sets indexed by $s$ such that $j$ and $j^\prime$ are in the same set $s$ if and only if they occur within one hour of each other.  With this convention, the marginalization over light-curve phases becomes
\begin{equation}
p\left(\{f_j\} | \{\bar f_b\}, A\right) \propto \prod_s \int_0^1 d\phi_s \, \prod_{j \in s} \exp -\frac{1}{2}\left(\frac{ f_j - \bar f_{b_j} \left[ 1 + Ah(\phi_s)\right]}{\sigma_j}\right)^2.
  \label{pA}
\end{equation}
Numerically, the integral over $\phi$ can be executed as a sum over $\approx 20$ equally-spaced samples of $0\le\phi<1.$   Using Bayes' theorem and assuming uniform priors for $\bar f_b$ and for $0\le A < 1$, we can take
\begin{equation}
  p\left(\{\bar f_b\}, A | \{f_j\} \right)) \propto p\left(\{f_j\} | \{\bar f_b\}, A\right).
\label{pbayes}
\end{equation}
We created a straightforward Metropolis-Hastings Markov Chain (MHMC) to sample values of $\bar f$ and $A$ from this posterior probability distribution. We can create samples from the posterior distribution of a TNO color $b-b^\prime$ by calculating $-2.5\log_{10}\bar f_b/\bar f_{b^\prime}$ for each step of the chain.  This will fully capture the (often non-Gaussian) distribution of the color.  Discarding the values of $A$ in the chain is equivalent to marginalizing over variability---or one can apply a prior on $A$ by weighting the color values by $p(A)$. 
Discarding the fluxes yields the posterior distribution of the LCA values $A$, which is also often non-Gaussian, whenever $A$ is poorly constrained.

\begin{figure}[ht!]
	\centering
	\includegraphics[width=0.75\textwidth]{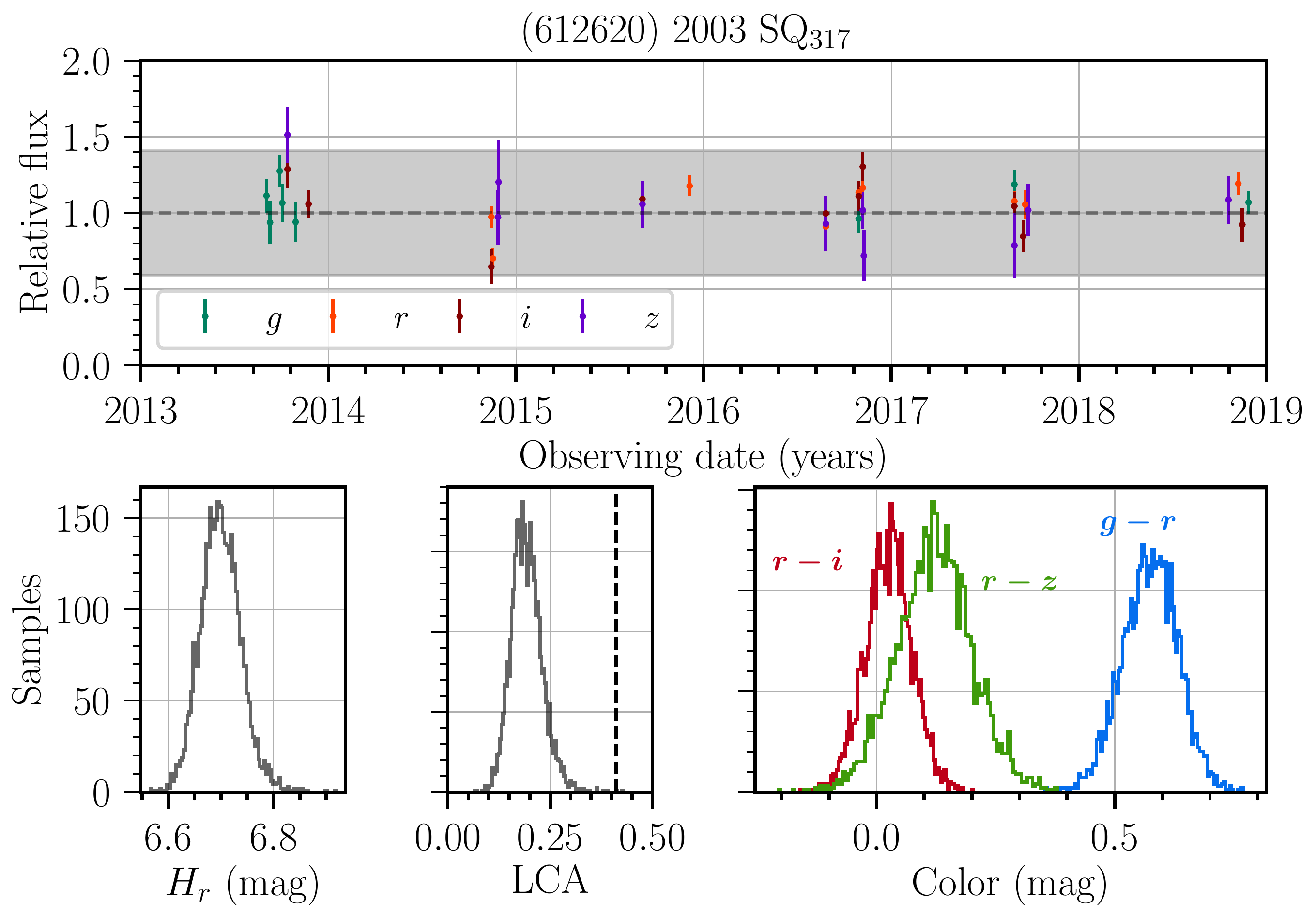}
	\vspace{-0.1cm}
	\includegraphics[width=0.75\textwidth]{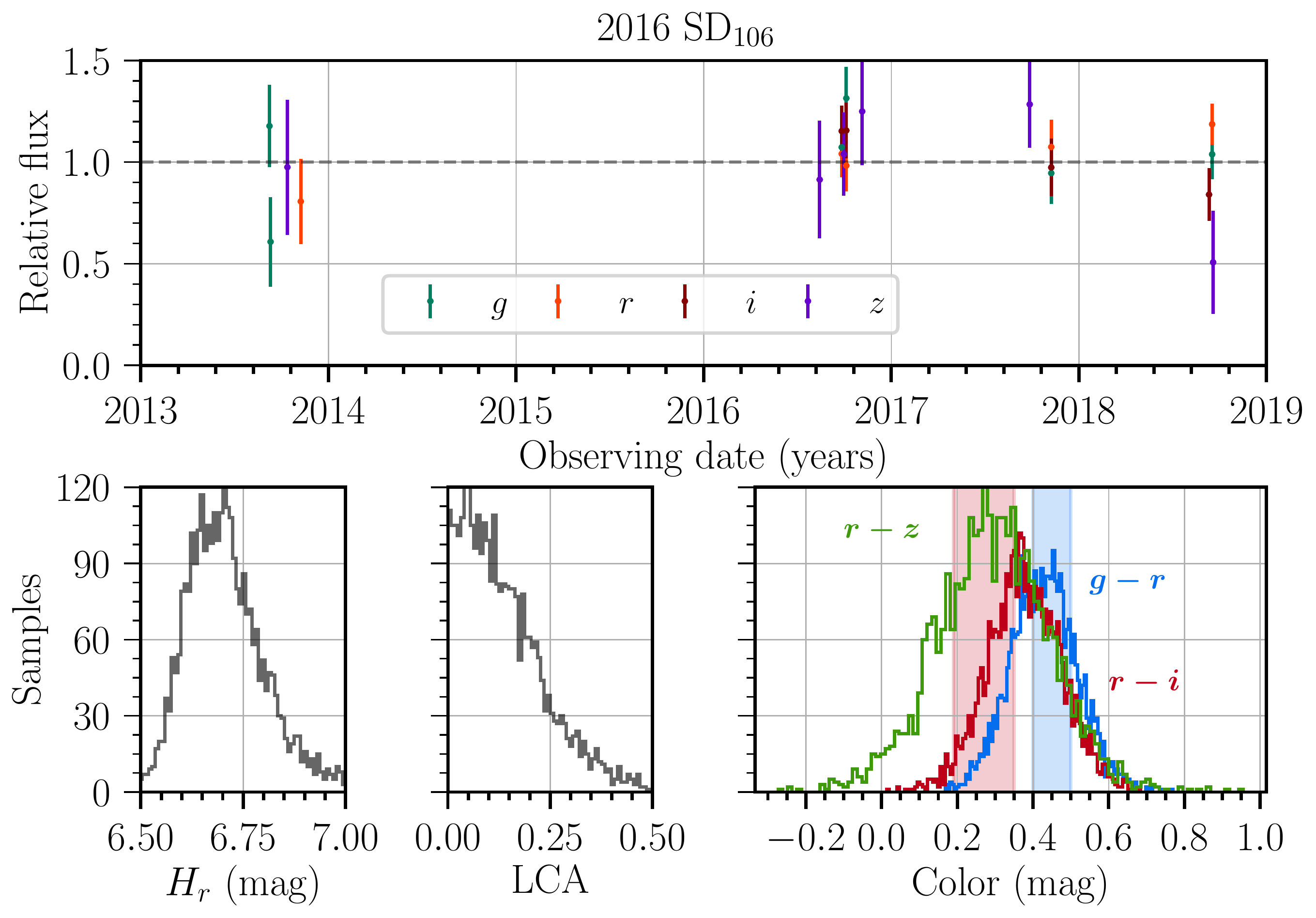}
	\vspace{-0.1cm}
	\caption{{The results of the modeling of the light curve amplitude and determination of mean flux and color are summarized for (612620) 2003 SQ$_{317}$ (top), a source with well-characterized variability;  and 2016 SD$_{106}$ (bottom), a source consistent with no variability.  For each source, the upper panel plots the measured fluxes over the six years of the survey, in particular the flux variation $f/\overline{f}_b$ relative to the mean flux in each band. The bottom panels for each object shows histograms of the MCMC samples drawing from the inferred posterior probabilities of 
the absolute magnitude $H_r$ (left), the light curve amplitude (LCA) $A$ (middle), and the colors (right) for the object. The horizontal band in the relative-flux figure and the dashed line in the LCA figure for (612620) 2003 SQ$_{317}$ show the light curve peak-to-peak amplitude measured by \cite{Lacerda2014}. Our method underestimates the LCA, likely because the true light curve is not sinusoidal. The vertical bands in the 2016 SD$_{106}$ color histograms show the 68\% limits from \cite{chen2022}, which are in good agreement with our inferred colors.} \label{im:phot}}
\end{figure}

The accuracy of the inferences on $A$ depend critically on having accurate estimates of the measurement errors $\sigma_j$ on individual epochs.  The careful characterization of image noise \citep{Bernstein2018} and photometric calibration \citep{Burke2017} of the \des\ data work together with the scene-modeling methods of Section~\ref{sec:smp} to return reliable uncertainties.
We take several further steps to guard against spurious measurements that will inflate the LCA estimate. First, every exposure of every TNO is visually inspected and those with image defects (bad columns, cosmic rays, scattered light, etc.) or poor scene-modelling residuals are excluded from the analysis. After the MHMC runs, any individual photometric data points that lie $>3\sigma$ outside of the span of the model light curve at the median $A$ value are clipped, and the MHMC is re-run.  We also visually inspect diagnostic plots of the MHMC (as in Figures~\ref{im:phot}) for outlying measurements or unusual posterior distributions, which results in identification of a handful of 
additional measurement issues. Figure \ref{im:phot} illustrates this procedure for two objects.

One validation of the LCA estimation is that the brightest observed TNO, Eris, returns a value of $A$ sharply constrained to a 68\% confidence interval of 0.020--0.025.  \citet{bernstein2023} determine the period and phased light curve of Eris from the combination of \des\ photometry with data from 3 other observing campaigns at different observatories, and derive a sinusoidal light curve with an amplitude of $A=0.015\pm0.001.$  The effect of spurious sources of fluctuation---such as calibration errors, illumination phase variation, photometric systematic errors, under-estimation of measurement errors---is to induce perhaps an extra $0.01$~mag of perceived variability semi-amplitude.  

{Another cross-check on the determination of the light-curve amplitude is comparison of our inferred $A\approx 0.2$ for  (612620) 2003 SQ$_{317}$ (implying $\Delta m=0.44$~mag) with the $\approx0.85$~mag peak-to-peak variability reported by \citet{Lacerda2014}.   As illustrated the topmost panel of Figure~\ref{im:phot}, it is plausible that \des\ photometry sampled variations in flux as broad as \citet{Lacerda2014} observed.  But the center panel of the second row shows that our estimated $A$ is incompatible with a sinusoidal light curve with $\Delta m=0.85$~mag.  This discrepancy is at least partially attributable to the fact that 2003 SQ$_{317}$ has a very non-sinusoidal light curve (as is characteristic of contact binaries).  It is also possible that the light curve is varying over time as the system precesses or our viewing angle changes.  This object has a rich long-term photometric dataset and is worthy of further investigation.  In any case, the discrepancy between our $A$ inference and the higher observed $\Delta m$ does not invalidate our methodology.  We are not using the sinusoidal-assumption $A$ as a definitive measure of peak-to-peak variability; we are using it as an indicative measure of variability that we can use to compare different TNO populations.  As long as we use a repeatable, well-defined measure of flux variability, such comparisons remain valid.}

\section{Constraining population-variance models}
\label{sec:lca}
\subsection{Probability calculation}
To make physical inferences about some selected subpopulation of TNOs, we will assume that their LCA's are drawn independently from some distribution $q(A | \theta_A),$ where $\theta_A$ is one or more parameter(s) characterizing the distribution.  Physically, the observed $q(A)$ will be some convolution of a distribution of intrinsic shapes (or surface variations) with a distribution of obliquity angles of the rotation axes to the line of sight.  An excellent overview of the derivation of these distributions is in \citet{Showalter2021}.  Previous analyses of TNO variability have,  however, been severely hampered by selection effects in the published values of LCA's---selections both on which objects were targeted, and on which were published.  Non-detections of variation or periods often go unpublished, and even published upper limits on LCA's have not been incorporated into population analyses in a rigorous way.  An advantage of the \des\ TNO sample is that we have multi-epoch photometry for \emph{all} of the targets, and we also know that the discovery rate is essentially independent of LCA \citep{Bernardinelli2022}, so there are no LCA-dependent selection effects.

Once a selection on orbital characteristics or $H$ is made, we have the posterior distribution $p(A_i|\{f_{ij}\})$ of source $i$ available in the form of samples from its MHMC chain, whether or not this posterior indicates a clear detection of variability.  We can now form a posterior likelihood for the distribution of the parameters $\theta_A$ of the LCA distribution for this class as
\begin{align}
  p(\theta_A | D) & \propto p(D | \theta_A) p(\theta_A)  \\
                  & = p(\theta_A) \prod_i p(\{f_{ij}\} | \theta_A) \\
                  & = p(\theta_A) \prod_i \int dA_i \,  p(\{f_{ij}\} | A_i) q(A_i | \theta_A) \\
                  & \propto  p(\theta_A) \prod_i \left[ \frac{1}{N_{{\rm samp},i}} \sum_{k=1}^{N_{{\rm samp},i}} q(A_{ik} | \theta_A) \right].
\label{qAsamp}
\end{align}
This follows from the use of Bayes' Theorem, plus the fact that the samples $A_{ik}$ from the MHMC chain of object $i$ are drawn from a distribution assuming a flat prior on $A_i.$  We can thus use \eqq{qAsamp} to assign a likelihood to any postulated LCA distribution for a population by summing over all the MHMC outputs, regardless of whether they indicate a decisive detection of variability.

In a similar vein, we can derive an estimate of the posterior distribution of color for any individual object by applying a chosen prior  $q(A_i | \theta_A)$ to the MHMC outputs, \eg\
\begin{equation}
  p(b-b^\prime | \{f_{ik}\}, \theta_A) \propto \sum_{k=1}^{N_{{\rm samp},i}} q(A_{ik} | \theta_A)\delta (b-b^\prime+2.5\log_{10}\bar f_{ibk}/\bar f_{ib^\prime k}).
\end{equation}

\subsection{Application to the \des\ TNOs}
One could in principle attempt to constrain the physical parameters of the TNO subpopulations by propagating such parameters through to a functional form for $q.$  Such modelling involves specifying the geometry of the objects, their surface scattering properties, and distribution of rotation axes distributions, as described in detail by \citet{Showalter2021}.  This is before one even considers the nature of albedo variations across the objects surface.  The available information is far short of what would be necessary to constrain the large number of free parameters in any realistic physical model.

We opt instead to adopt a simple heuristic parametric form for $q(A),$ which we will fit to various subpopulations with the intention of looking at the distinctions across the trans-Neptunian region, rather than attempting to extract physical parameters.  In other words we will use $q(A)$ as a kind of genetic marker for the relationships among the subpopulations.

A flexible family of distributions normalized over the range $0\le A<1$ is the $\beta$ distribution,  with a probability distribution function usually written as
\begin{equation}
  q(A | a,b) \propto A^{a-1} (1-A)^{b-1}.
\end{equation}
We transform the parameters slightly, using first the mean of the distribution, $\bar A = a/(a+b),$ and a second parameter $s\equiv \log_{10}(a+b),$ a ``sharpness'' parameter: at fixed $\bar A,$ the $\beta$ distribution becomes narrower about the mean as $s$ increases.  Thus we will have
\begin{equation}
  q(A | \theta_A) = q(A | \bar A, s) \propto A^{\bar A 10^s -1} (1-A)^{(1-\bar A)10^s-1}.
  \label{eq:qa}.
\end{equation}
The $\beta$ distributions are a superset of the power-law distributions considered in early modeling of the variability of TNOs \citep{Lacerda2006}.
Since \citet{Bernardinelli2022} show that the selection function of the \des\ survey is nearly independent of $A$ at fixed mean apparent magnitude, we do not need to introduce selection terms into our posterior probability for the parameters of $q(A).$

For a chosen sample of \des\ TNOs, we calculate the posterior probability of $(\bar A,s)$ using Bayes' Theorem, $p(\bar A, s | D) \propto p(D | \bar A, s) p(\bar A, s),$ where the data $D$ consist of the MCMC chains sampling the $A_i$ distribution for TNO $i$ within the sample.  This can be evaluated across the full $(\bar A, s)$ space numerically using \eqq{qAsamp}, and assuming uniform priors over $\bar A$ and $s.$
We will also marginalize over $s$ to obtain $p(\bar A | D)$ for different subpopulations.  The overlap between these distributions for two different subpopulations becomes a measure of their morphological similarity.

\begin{figure}
\centering
\includegraphics[width=0.7\textwidth]{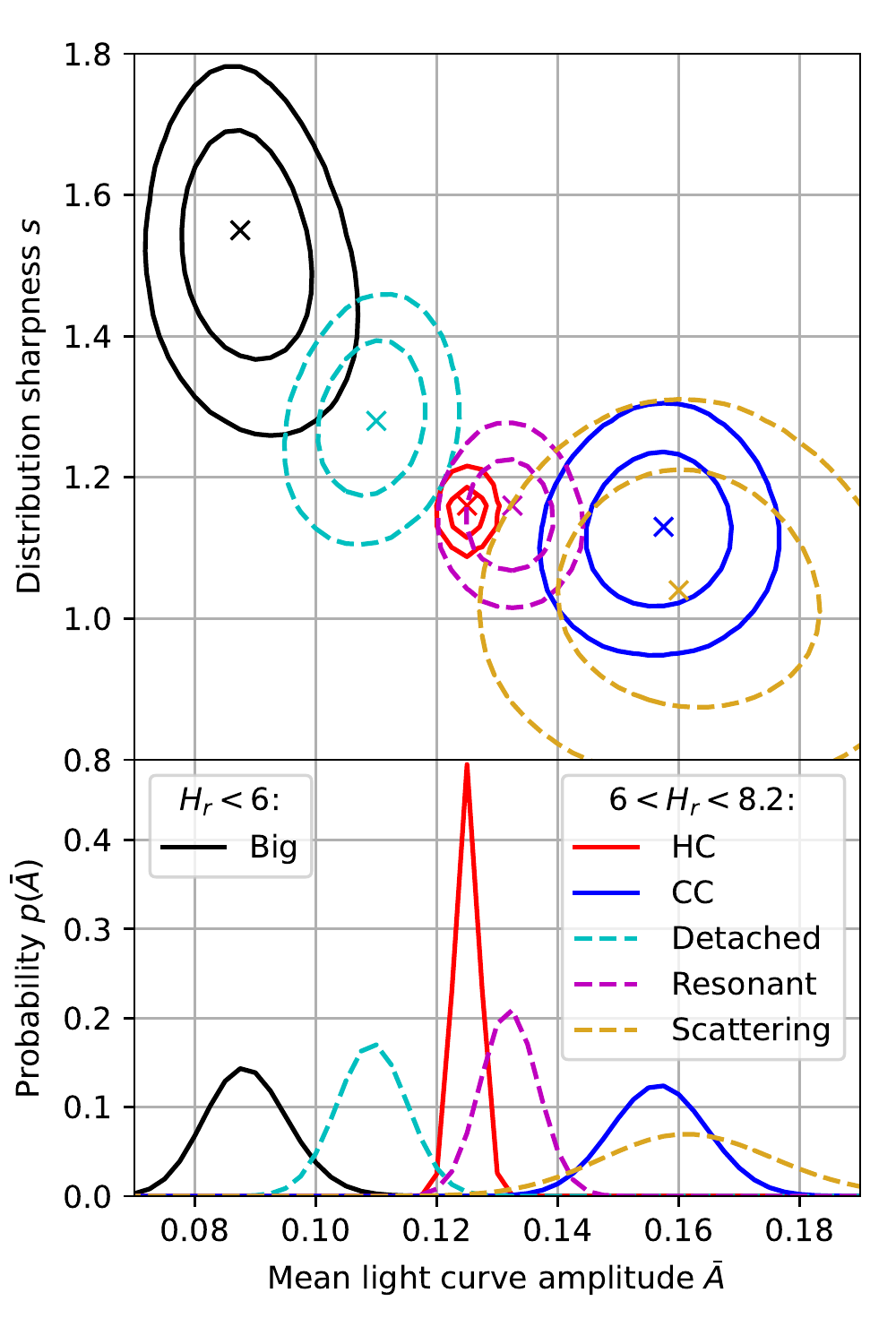}
\caption{For each of six disjoint subsamples of TNOs described in Table~\ref{samples}, the top plot shows their 68\% and 95\% posterior credible regions of the $(\bar A, s)$ distribution parameters as used in \eqq{eq:qa}.  The maximum posterior likelihood of each is marked with an ``X.''  The bottom panel projects these posterior distributions onto the single parameter $\bar A$ that gives the mean of the inferred distribution of noiseless light curve amplitudes for that population.
It is immediately clear that larger TNOs (``Big'') with $H_r<6$ are less variable than the smaller members of all the dynamical populations, with the possible exception of detached TNOs.  Furthermore the hot classicals (``HC'') are less variable than the cold classicals (``CC''), with the resonant population being consistent with either and our constraints on the scattering population being weaker. Detached TNOs are also decisively less variable than resonant or scattering TNOs in the same $H_r$ range.}
\label{im:abar}
\end{figure}

Table~\ref{samples} defines several disjoint samples for which we calculate $p(\bar A, s)$, and Figure~\ref{im:abar} plots the posterior probability in the $(\bar A, s)$ space, and marginalized down to $p(\bar A)$ for each.  It is expected that more massive TNOs will be more spherical due to self-gravity---and hence have $q(A)$ distributions weighted toward lower values---so it is important to control for size when testing differences.  All but one of the TNO samples we define therefore are restricted to $6<H_r<8.2$ (and have $\langle H_r \rangle=7.4\pm 0.2),$ a range in which most of the dynamical families have a useful fraction of their \des-detected TNOs. .  The ``Big'' sample contains all \des\ TNOs with $H_r<6.$  Aside from Eris, this sample's members have $3.3<H_r<6.0,$ and only two of them meet the cold-classical criterion, so the ``Big'' subset is essentially composed of dynamically excited TNOs.

\begin{deluxetable}{lrcccccl}
\tablewidth{0pt}
\tablecaption{TNO sample definitions}
\tabletypesize{\small}
  \tablehead{\colhead{Subset} & \colhead{$N$} & \colhead{Size} & \colhead{Median$(H_r)$} & \colhead{Dynamical Class} & \colhead{Inclination} 
  & \colhead{$\bar A$ 68\% CL}}
\startdata
Cold classical (CC) & 95 & $6<H_r<8.2$ & 7.18 & Classical & $i_{\rm free}<5^\circ$ & 0.149--0.165 \\
Hot Classical (HC) & 261 & $6<H_r<8.2$ & 7.38 & Classical & $i_{\rm free}>5^\circ$ &  0.124--0.128 \\
Resonant & 170 & $6<H_r<8.2$ & 7.56 & Resonant & \nodata &  0.127--0.136 \\
Detached & 133 & $6<H_r<8.2$ & 7.39 & Detached & \nodata &  0.114--0.115 \\
Scattering & 34 & $6<H_r<8.2$ & 7.57 & Scattering & \nodata &  0.148--0.176 \\
Big & 36 & $H_r < 6$ & 5.43 & \nodata & \nodata &  0.081--0.095
\enddata
\tablecomments{Each row gives the conditions defining one subsample of the \des\ TNO sample, and the resultant number and median absolute magnitude $H_r$ of the sample.  Dynamical classification are taken from \citet{Bernardinelli2022} and the free inclinations $i_\mathrm{free}$ of all classical objects are provided by \citet{Huang2022}. The final column gives the 68\% confidence interval for the mean light curve amplitude $\bar A$ of the distribution of the subset's members' $A$ values.}
\label{samples}
\end{deluxetable}

{Figure~\ref{im:abar}} immediately indicates that the ``Big'' $H_r<6$ sample is indeed less variable {(lower $\bar A$)} than the $6<H_r<8.2$ members of any of the dynamical classes, as might be expected from the effects of self-gravity.  We wish to determine whether two subsamples $D_A$ and $D_B$ could be drawn from the same parent distribution of parameters.  Hypothesis $H_1$ is that they are from a single $\beta$ distribution.  Hypothesis $H_2$ is that they are from distinct distributions with two pairs of $\beta$-distribution parameters.  We give two ways of determining the strength of a conclusion in favor of $H_2,$ distinct distributions.  A frequentist-oriented statistic is
\begin{equation}
  \Delta\chi^2 \equiv 2\log \frac{ \left[\max_{\bar A, s} p(\bar A, s | D_A)\right] \left[\max_{\bar A, s}  p(\bar A, s | D_B)\right]}
  {\max_{\bar A, s}  p(\bar A, s | D_A,D_B)}.
\label{eq:dchisq}
 \end{equation}
 This statistic compares the maximum posterior likelihood of $H_2$ to $H_1.$  In some conditions (which we do not actually meet), this statistic would have a $\chi^2$ distribution with $\nu=2$ degrees of freedom, if the two populations were indeed drawn from the same model. Then the probability of $\Delta\chi^2 > 4.6$ would be 10\% and $\Delta \chi^2 > 9.2$ would be 1\%.  We will designate these as ``indicative'' and ``decisive'' evidence of differences in the underlying $q(A)$ distribution of two samples.  By this criterion, the ``Big'' sample is indeed decisively distinct from all the higher-$H_r$ subpopulations, except for being indicative for the detached TNOs.

 A more Bayesian statistic is the (log) evidence ratio, defined as
 \begin{equation}
   {\cal R} = \log \frac{p(D | H_2)}{p(D | H_1)} = \log \frac{ \int d\bar A\, ds \, p(D_A|\bar A, s) p(\bar A,s)\times  \int d\bar A\, ds \, p(D_B|\bar A, s) p(\bar A,s) }{\int d\bar A\, ds \, p(D_B,D_A|\bar A, s) p(\bar A,s) }.
   \label{eq:evidence}
 \end{equation}
 Adopting the scale of \citet{jeffreys}, values of $\mathcal{R}>2.30,$ 3.45, and 4.61 are labelled as ``strong,'' ``very strong,'' and ``decisive'' in favor of $H_2.$  The first and last of these correspond to $\approx10\%$ and $\approx1\%$ chances of $H_1$ being correct, if we assign equal prior probability to $H_1$ and $H_2.$

For our problem, we require a normalized prior $p(\bar A, s).$ We choose a prior which is uniform over the range $0.06<\bar A<0.19,$ $0.7<s<1.9.$ 
The evidence ratio is indeed decisively in favor of the Big size distribution being distinct from all the other subsets, except the detached population, for which ${\cal R}=0.54$ is indecisive about $H_1$ vs $H_2.$

Also conclusive from the $\Delta\chi^2$ test and ``very strong'' at ${\cal R}=4.33$ in the evidence ratio is the hypothesis that the cold classicals at $i_{\rm free}<5^\circ$ are distinct, namely more variable, than the hot classicals ($i_{\rm free}>5^\circ$). 
The CC's are believed to be the most dynamically pristine, and are known to have redder colors and a higher rate of binarity \citep{Stephens2006,Fraser2012,NESVORNY201949}.
Our results demonstrate that variability is another physical distinction between higher- and lower-inclination classical TNOs.  

\begin{figure}
  \centering
  \includegraphics[width=0.45\textwidth]{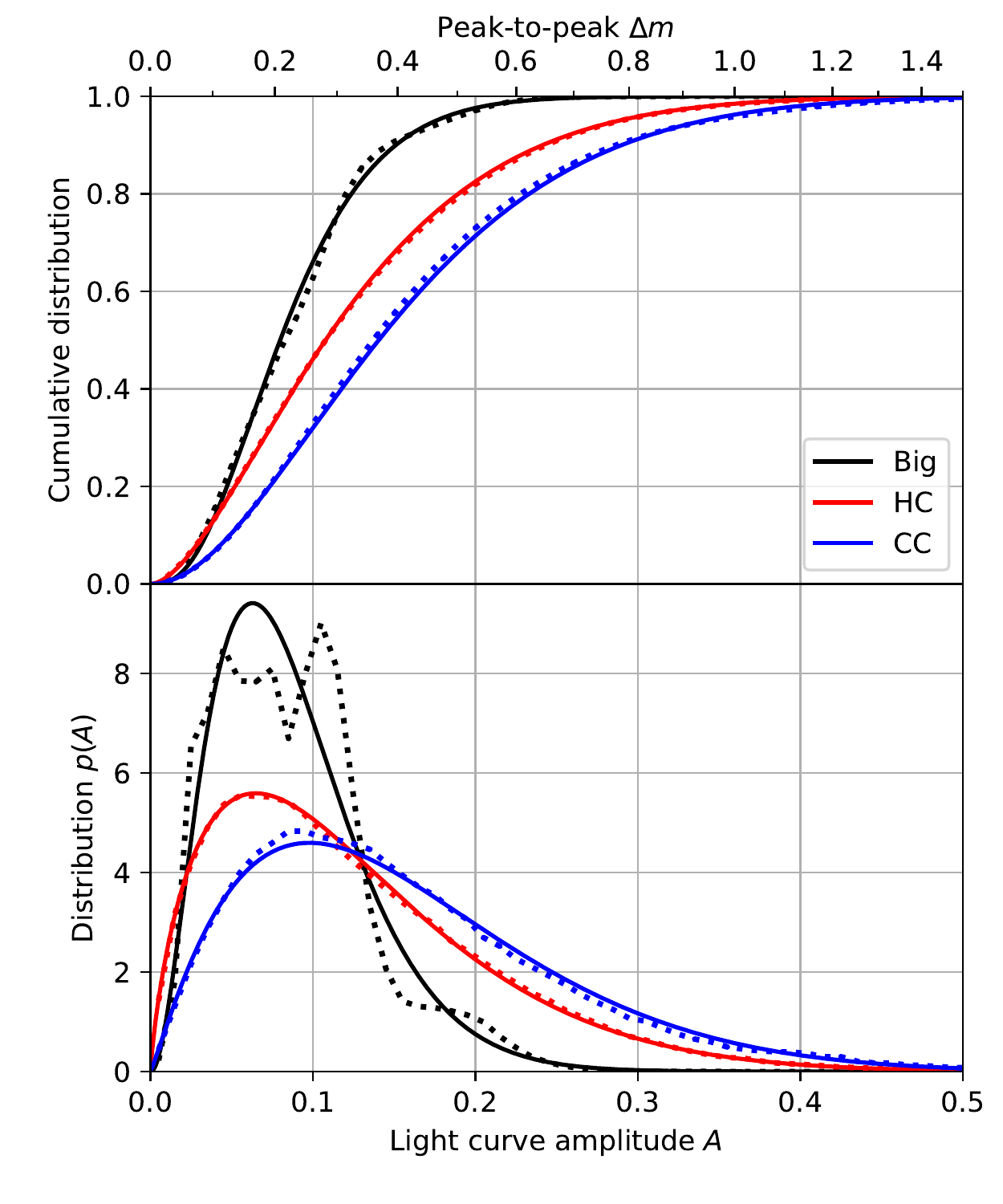}
  \includegraphics[width=0.45\textwidth]{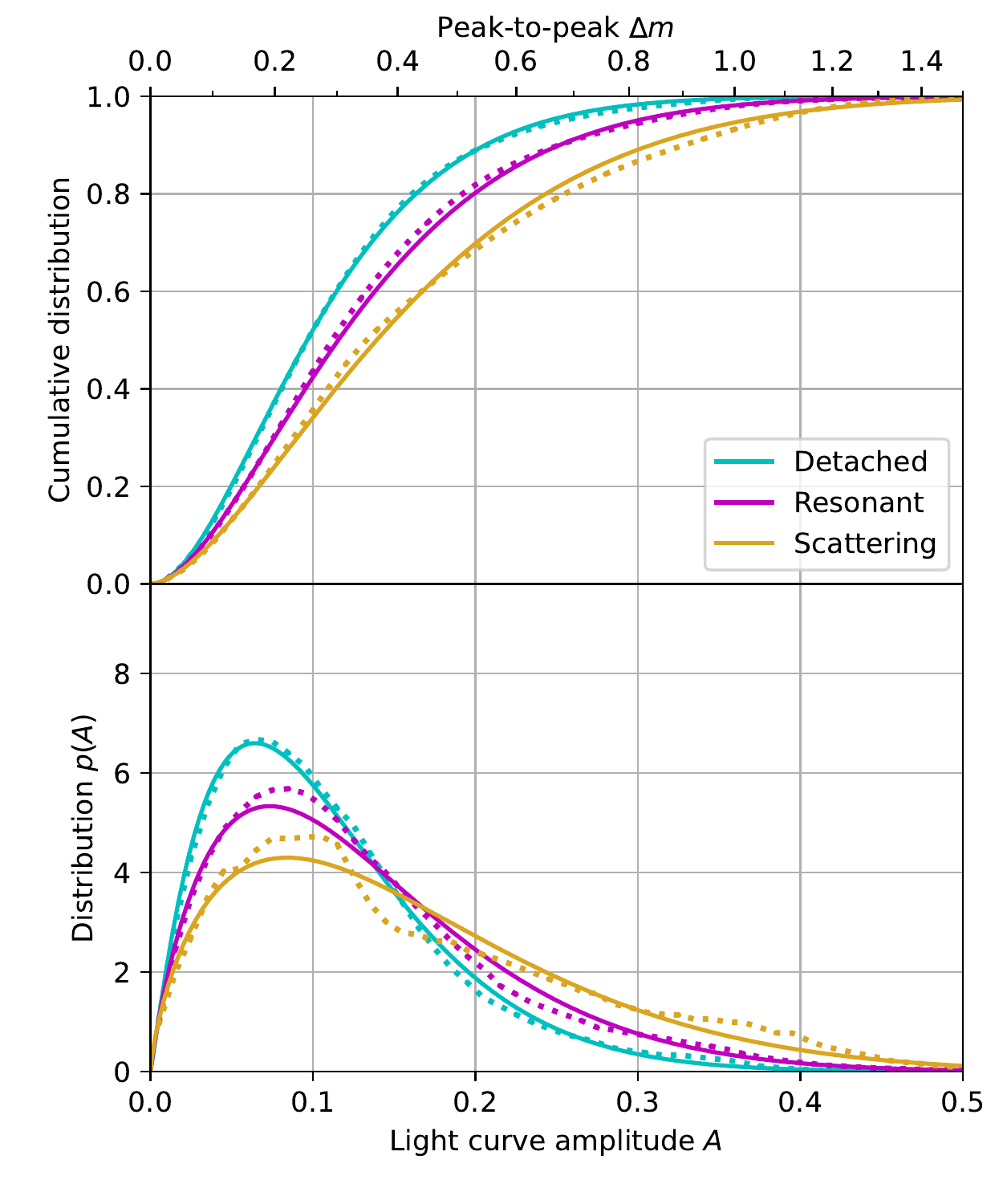}
  \caption{The best-fit $\beta$-distributions $q(A)$ of the light-curve amplitude for the marked subsets are plotted in the lower panels as solid lines, and the corresponding cumulative distributions in the upper panels.  The dotted lines are the posterior distributions for $A$ of the observed sample when using these best-fit functions as priors.  The agreement between the data and the $\beta$ distributions are good, with larger deviations seen (as expected) in the samples with fewer members.  The top axis converts our sinusoidal flux variation amplitude $A$ into the peak-to-peak magnitude change of the light curve (under the sinusoidal assumption), for easier comparison to the literature.  The upper axis only applies to the upper panels.
}
  \label{im:bestfit}
\end{figure}

Before proceeding further, we check whether the $\beta$ distributions are adequate descriptions of the measured size distributions.  In the left panel of Figure~\ref{im:bestfit}, we plot in solid lines the $q(A)$ $\beta$-distributions that maximize the posterior probability for each of the Big, HC, and CC samples.  Both the differential and cumulative distributions are given.  The dotted curves plot the \emph{posterior} distributions for the observed data, \ie\ we average the posterior distributions $p(A|D_i)=p(D_i|A)q(A) / \int dA\, p(D_i|A)q(A)$ over all members of the labelled subset.\footnote{Note that this posterior distribution would equal the observed distribution if every TNO's value of $A$ were well constrained.}  The agreement is good, with deviations that are larger for the subsets with fewer members.

Next we examine the size-controlled samples of the other dynamically excited populations, the resonant, scattering, and detached TNOs.
Neither the resonant and scattered TNO populations can be confidently distinguished from the HC's or CC's in terms of their $q(A)$ parameters.  Indeed the evidence ratio ${\cal R}=-3.53$ for the HC-resonant pair indicates very strong preference for hypothesis of a shared distribution.  By the $\Delta\chi^2$ criterion, there is indicative evidence that the resonants are distinct from the CCs, and that the scattering TNOs are distinct from the HC's.  We can summarize this by saying that the variabilities of the resonant and scattered populations are fully consistent with being somewhere between the HC's and CC's.

The last dynamically excited population, the detached TNOs, lies between the HC and Big loci, and is decisively distinct from CC's in variability by all criteria.  The evidence ratio indicates that the detached TNOs could be from the same distribution as either the Big or HC, while the $\Delta\chi^2$ offers indicative and decisive distinction from the Big and HC populations, respectively.  A surprising result is that the detached and scattering TNOs are decisively inconsistent by both criteria.

The results of both forms of model comparison test can be easily summarized: if the 95\% CL ellipses of two populations do not overlap in
in the upper panel of Figure~\ref{im:abar}, then they are decisively distinct from each other by one or both of the more formal criteria.

\subsection{Discussion}
Further division of the \des\ TNOs into smaller subsets than listed in Table~\ref{samples} did not yield any ability to distinguish statistically significant differences.  In particular,
we have tested for dependence on $H_r$ within the HC sample by splitting at the median $H_r;$ neither hypothesis test indicates a preference for distinct $q(A)$ between the halves.
Similarly we split the resonant class into high-$a$ and low-$a$ halves; high-$i$ and low-$i$ halves; and Plutinos vs others.  In no case did the splits yield significantly better modeling.  Of course this does not preclude the existence of such dependence, but the quantity and quality of our data are insufficient to detect any.

Comparison of these conclusions with previous investigations, and implications for the origin of the populations, are discussed in Section~\ref{sec:conclusions}.

\section{Conclusions}
\label{sec:conclusions}
\subsection{Methodologies}
This paper describes a series of techniques to obtain optimal measures of fluxes, colors, binarity, and variability of moving sources like TNOs, with principled estimation of uncertainties.   We apply these techniques to the $>800$ TNOs detected by the \des, yielding accurate colors for each, discovery of several likely new wide binaries, and the ability to make inferences about the relative levels of variability of TNO populations with more rigour and confidence than previous studies of TNO variability.  Aside from the $\approx10\times$ larger sample of TNOs than preceding studies, this analysis addresses several methodological issues that affect some investigations:
\begin{itemize}
\item The photometry methods obtain the optimal signal-to-noise level on TNO fluxes from every frame by using PSF-fitting techniques, and also subtract any background sources that may be behind the TNO---even those that may be undetectable at the noise level of the target image. 
\item The estimates of mean $H_r$ and colors make use of \emph{all} of the exposures of a given TNO, avoiding biases that would result from using only the images with $S/N$ sufficient for detection.  The color estimates and their uncertainties include the effects of potential variability of the source while exploiting any information from near-simultaneous colors.
\item Estimates of the variability yield a full probability distribution for $A,$ rather than arbitrary decisions about and treatment of non-detections.  This is true whether or not a period is detectable.  This enables a correct quantitative comparison to candidate distributions $q(A),$ even for TNOs with poorly constrained $A.$
\item The analysis includes \emph{all} of the TNOs detected by \des\ that meet selection criteria on dynamics or $H.$  In other words the selection function for the input sample is well defined and independent of variability.  Most previous studies have relied on samples drawn from the very heterogeneous literature, which are subject to unquantifiable selection biases.  Most pernicious might be a publication bias, whereby many authors might only publish amplitudes of various if they have detected variation, leaving out non-detections.
\item The 3--5 year temporal spread of the \des\ photometry is far longer than any TNO rotation period, insuring that light-curve phases are randomly sampled and our variability estimates are independent of the period.
  \item The \des\ sample is large enough to make meaningful comparisons between dynamical groups at a fixed and limited range of $H$ so we can separate dynamical-state dependence from size dependence.
  \end{itemize}

  These techniques should be of great value for future, even larger-scale sky surveys, such as the \textit{Legacy Survey of Space and Time,} which will measure at least an order of magnitude more TNOs than \des\ has found.  The vast majority of TNO measurements in any survey have $S/N$ levels too low to allow traditional measurements of light curves, but there is still a great deal of information available from their colors and variability.

\subsection{Variability distributions}
The level of variability in our ``Big'' sample ($3.3<H_r<6.0,$ plus Eris) is decisively lower than that any of the dynamical classes' $6.0<H_r<8.2$ members.  A relative paucity of variability at $\Delta m >0.3$~mag among $H\lesssim 6$ objects has been noted in past works, and is readily ascribed to the effects of greater surface gravity in forcing objects toward sphericity.  \citet{Benecchi2013}, for example, find a correlation of light-curve amplitude vs $H_V$  at $3\sigma$ significance in a sample comprised of 32 objects at $H_V<6$ that they measure, combined with 96 objects at $0<H_V<12$ with $\Delta m$ measures drawn from the extant literature, mixing together all dynamical classes. As noted by \citet{Showalter2021} in reviewing variability statistics, we should be alert to selection biases and publication biases in samples drawn from the literature.  They opt to nonetheless ``proceed on the assumption that our sample is unbiased simply because we have no clear alternative.''  With the \des\ sample we no longer need to make this leap.
The measurements by
\citet{OSSOSvariability} also have a selection criterion that should be independent of $A,$ since they target all 63 OSSOS TNO detections at $6<H_r<11$ within a few chosen fields of view of the HyperSuprimeCam imager.  Within this sample they report a correlation between $\sigma_{\rm mag}$ (the observed magnitude dispersion per object) with $H_r$ at $p=0.013$ false-positive rate in a Spearman rank correlation test.  The correlation persists at $p=0.015$ when the sample is restricted to dynamically excited populations (\ie\ excluding CCs).\footnote{We focus on the $\sigma_{\rm mag}$ statistics from \citet{OSSOSvariability}  because it is less susceptible to measurement noise and outliers, such as from background sources, than their $\Delta_{\rm mag}$ statistic of extremal measurements.  \citet{ThirouinCC} suggest that some of the outlying OSSOS photometry needs to be clipped, and the methodology also has no allowance for removal of measurement noise. \citet{Showalter2021} find the OSSOS distributions discrepant from other data, but the other data come from ill-defined samples.}  Our data can put to rest any lingering doubt about the reality of a variability-$H$ connection, since we use samples guaranteed to be free of variability biases, and can show that the $3.3<H_r<6$ sample is distinctly less variable from the smaller objects in every dynamical class (with the possible exception of detached TNOs). Our distributions for $H_r \leq 6$ is also in general agreement with the targeted sample of primarily large objects by \cite{Kecskemethy2023}, who show that $\approx 55\%$ of their sample has $\Delta m \leq 0.2$ mag (compare to the upper-left panel of Figure~\ref{im:bestfit}).

Another decisive result of our variability study is a higher level of variability in ``cold'' classical TNOs ($i_{\rm free}<5^\circ$) than in ``hot'' classicals ($i_{\rm free}>5^\circ$).  Correlations of variability with inclination, or higher variability in CC's than in non-CC's, have been noted in other samples.
\citet{Benecchi2013} report differences at $\approx95\%$ CL between the variabilities of classical TNOs and those of resonant and/or scattered objects, and detect a correlation between $\Delta m$ and $i$ at similar confidence across the full TNO sample.  The decisive difference between HC and CC populations' varibilities seen in our data reinforce the suspicions based on dynamical and color differences that the CC's have a distinct physical origin, and we should therefore split these in any comparison to other dynamical families.  We should also realize that the simple $i_{\rm free}$ cut is probably not the definitive division between these physical two populations, so our HC and CC samples are a bit mixed, which means we are potentially underestimating their distinction.

\citet{ThirouinCC} have compared their own measures of CC variability against other populations' data in the literature, reporting that 36\% of CC's have $\Delta m<0.2$~mag---quite close to our value on the blue curve in the upper left of Figure~\ref{im:bestfit}---compared to 65\% of a mixed bag of ``other'' TNOs.  Our data agree with this trend in the sense of CCs being the most variable of all the dynamical classes, at fixed $H$ range. \citet{OSSOSvariability}  also report a correlation between $i$ and $H_r$ \emph{within} the CC class at 94\% CL.  We do not confirm this trend (nor falsify it): splitting the CC sample at the median $i_{\rm free}$ does not yield statistically significant distinctions in $q(A).$  Neither does splitting the much larger HC sample at its median $i_{\rm free}$ of $15^\circ.$  Our data support only the dichotomy of variability between HC and CC dynamical types, no further evidence exists for a continuous gradient in $i_{\rm free}.$

The resonant and scattering TNOs are not distinguishable from either the HC or CC TNOs with regards to their variability levels.  This is consistent with any scenario in which resonant or scattering TNOs share a physical origin with any mixture of the classicals' source populations.  The surprising result, however, is that the detached and scattering populations have decisively distinct variability distributions ($\Delta \chi^2=18.2, {\mathcal R}=6.5$).  The detached and resonant populations meet the frequentist criterion for decisive distinction ($\Delta \chi^2=12.0, {\mathcal R}=2.4$).  This would exclude scenarios in which members of the scattering population have their perihelia raised and become detached by a process that is independent of the physical nature of the TNOs.
This might be consistent with the observations if TNOs were substantially physically altered in the detaching process, to lower their mean variability by a factor $\approx 1.5$.  
An isotropization of spin axes that were initially aligned normal to the ecliptic could lower the mean apparent variability by this much, but one would need to explain why the detached TNOs are isotropized but the scattered TNOs are not. A weaker version of this restriction applies to the detached TNOs arising through the resonant channel.  This is a puzzle for models of the origin of the detached TNOs. 

\subsection{Binary discoveries}
We discover one new binary, in addition to the previously known 2014 LQ$_{28}$, and characterize their colors and sizes. Our ability to identify these binaries are a proof-of-concept of the combination of scene modeling photometry with binary deblending. These objects also have enough astrometric data over the sparse \des\ observations that we can determine a mutual orbit for the binary system. All of our objects are tightly bound, with $a_m/R_H \approx 5\%$. Their masses are in significant agreement with masses of other objects of similar sizes reported in the literature \citep[\emph{e.g.}][]{GRUNDY2011678,Parker2011a}.

 2014 LQ$_{28}$ and 2013 RJ$_{124}$ are cold Classicals, adding to the now large sample of characterized binary CCs \citep{2020tnss.book..201N}. Notably, 2014 LQ$_{28}$ is the second largest CC in the \des\ sample ($H_r \approx 5.7$), which is in line with the hypothesis that most large CCs are binaries \citep{thirouin2019}. We can estimate from the differences in $H_r$ and masses that 2013 RJ$_{124}$ is either half as dense or half as reflective as 2014 LQ$_{28}$. Assuming the same albedo for both objects, the volume ratios between both objects is $2.8$, while the mass ratio is $\approx 1.38$. On the other hand, if we assume that 2014 LQ$_{28}$'s albedo is twice that of 2013 RJ$_{124}$, we obtain roughly the same volume for both objects, and such albedo differences are in line with measurements from thermal data \citep{vilenius2012}.

Readers should resist the temptation to use our binarity detections to measure the fraction of wide binaries in the TNO populations. While the binary search was exhaustive over all our images, it was not systematically characterized, meaning that we have not characterized our ability to discover binaries from our images as a function of mutual orbit parameters, seeing, and $S/N$. The binary discovery fraction in the \des\ data strongly underestimates the binary fraction in TNO populations, since the majority of the objects detected in our survey are near the detection threshold, where our ability to resolve binaries is inherently diminished.  Further observations of each of these binary objects, with even slightly better angular resolution, would greatly improve constraints on their orbits and masses.

\subsection{Data release}
\label{sec:datarelease}
%\gary{Pedro, what do you think of this list?}

The data release presented here substantially increases the number of TNOs with known colors and LCAs in the literature, and is the largest of such kind from a single survey with a consistent filter set and calibration. 
Digital resources containing the results of application of these techniques to the $>800$ TNOs detected in the \des\ Wide Survey are available at [a site to be opened upon acceptance of this paper by the journal].  Included in the data release are:
\begin{itemize}
\item A table of all the valid individual photometric flux measurements of all the TNOs.
\item A table of all photometric measurements of the resolved binaries.
\item For the two TNO binaries we also include the MCMC chains with their derived mutual orbits.
\item A \textit{Jupyter} notebook containing code that can run MCMC chains to sample the space of mean fluxes and variability, given a set of multiband, sparsely sampled fluxes as described in Section~\ref{sec:colormc}.
\item For each TNO, a table giving the output MCMC chains produced from its photometry.
\item An update of the table of TNO properties given by \citet{Bernardinelli2022}, augmented with the means and standard deviations of $H_r$, colors, and light-curve amplitudes $A$ for each object, and derived from the MCMC chains.
\item A \textit{Jupyter} notebook that derives the plots and statistics of this paper from the MCMC chains, and demonstrates how a user could apply the methodology to other subsets of the \des\ TNOs or to other data.
\end{itemize}
\bibliography{references}
\bibliographystyle{aasjournal}
%\nocite{*}
\acknowledgments

\emph{Software}: This work made use of the following public codes: \textsc{Numpy} \citep{Numpy}, \textsc{SciPy} \citep{SciPy}, \textsc{Astropy} \citep{Astropy2013,Astropy2018}, \textsc{Matplotlib} \citep{Matplotlib}, \textsc{IPython} \citep{iPython}, \textsc{WCSFit} and \textsc{pixmappy} \citep{Bernstein2017}, \textsc{Piff} \citep{Jarvis2020}, \textsc{emcee} \citep{foreman-mackey2013}.

PHB acknowledges support from the DIRAC Institute in the Department of Astronomy at the University of Washington. The DIRAC Institute is supported through generous gifts from the Charles and Lisa Simonyi Fund for Arts and Sciences, and the Washington Research Foundation. Work by GMB, PHB, and NJ was supported by National Science Foundation grants AST-2009210 and AST-2205808.

Funding for the DES Projects has been provided by the U.S. Department of Energy, the U.S. National Science Foundation, the Ministry of Science and Education of Spain,
the Science and Technology Facilities Council of the United Kingdom, the Higher Education Funding Council for England, the National Center for Supercomputing
Applications at the University of Illinois at Urbana-Champaign, the Kavli Institute of Cosmological Physics at the University of Chicago,
the Center for Cosmology and Astro-Particle Physics at the Ohio State University,
the Mitchell Institute for Fundamental Physics and Astronomy at Texas A\&M University, Financiadora de Estudos e Projetos,
Funda{\c c}{\~a}o Carlos Chagas Filho de Amparo {\`a} Pesquisa do Estado do Rio de Janeiro, Conselho Nacional de Desenvolvimento Cient{\'i}fico e Tecnol{\'o}gico and
the Minist{\'e}rio da Ci{\^e}ncia, Tecnologia e Inova{\c c}{\~a}o, the Deutsche Forschungsgemeinschaft and the Collaborating Institutions in the Dark Energy Survey.

The Collaborating Institutions are Argonne National Laboratory, the University of California at Santa Cruz, the University of Cambridge, Centro de Investigaciones Energ{\'e}ticas,
Medioambientales y Tecnol{\'o}gicas-Madrid, the University of Chicago, University College London, the DES-Brazil Consortium, the University of Edinburgh,
the Eidgen{\"o}ssische Technische Hochschule (ETH) Z{\"u}rich,
Fermi National Accelerator Laboratory, the University of Illinois at Urbana-Champaign, the Institut de Ci{\`e}ncies de l'Espai (IEEC/CSIC),
the Institut de F{\'i}sica d'Altes Energies, Lawrence Berkeley National Laboratory, the Ludwig-Maximilians Universit{\"a}t M{\"u}nchen and the associated Excellence Cluster Universe,
the University of Michigan, the National Optical Astronomy Observatory, the University of Nottingham, The Ohio State University, the University of Pennsylvania, the University of Portsmouth,
SLAC National Accelerator Laboratory, Stanford University, the University of Sussex, Texas A\&M University, and the OzDES Membership Consortium.

Based in part on observations at Cerro Tololo Inter-American Observatory, National Optical-Infrared Astronomy Observatory, which is operated by the Association of 
Universities for Research in Astronomy (AURA) under a cooperative agreement with the National Science Foundation.

The DES data management system is supported by the National Science Foundation under Grant Numbers AST-1138766 and AST-1536171.
The DES participants from Spanish institutions are partially supported by MINECO under grants AYA2015-71825, ESP2015-66861, FPA2015-68048, SEV-2016-0588, SEV-2016-0597, and MDM-2015-0509,
some of which include ERDF funds from the European Union. IFAE is partially funded by the CERCA program of the Generalitat de Catalunya.
Research leading to these results has received funding from the European Research
Council under the European Union's Seventh Framework Program (FP7/2007-2013) including ERC grant agreements 240672, 291329, and 306478.
We acknowledge support from the Australian Research Council Centre of Excellence for All-sky Astrophysics (CAASTRO), through project number CE110001020, and the Brazilian Instituto Nacional de Ci\^encia
e Tecnologia (INCT) e-Universe (CNPq grant 465376/2014-2).

This manuscript has been authored by Fermi Research Alliance, LLC under Contract No. DE-AC02-07CH11359 with the U.S. Department of Energy, Office of Science, Office of High Energy Physics. The United States Government retains and the publisher, by accepting the article for publication, acknowledges that the United States Government retains a non-exclusive, paid-up, irrevocable, world-wide license to publish or reproduce the published form of this manuscript, or allow others to do so, for United States Government purposes.

\allauthors

\end{document}